\documentclass[11pt,a4paper]{article}

\usepackage{graphicx}
\usepackage{epsfig}
\usepackage[titletoc,toc,title]{appendix}

\usepackage{hyperref}
\usepackage{color}
\usepackage{ mathrsfs }

\usepackage{bm}
\usepackage{amsmath}

\newcommand{\vecb}[1]{{\bf #1}}
\def \d {\mathrm{d}}
\begin{document}

\begin{center}
{\LARGE{\textbf{Nonlinear interplay of Alfv\'en instabilities and energetic particles in tokamaks.}}}\\
\vspace{0.2 cm}
{\normalsize {\underline{A. Biancalani}${}^{1}$, A. Bottino${}^{1}$, M. Cole${}^{2}$, C. Di Troia${}^{3}$, Ph. Lauber${}^{1}$, A. Mishchenko${}^{2}$,\\
B. D. Scott${}^{1}$, F. Zonca${}^{3,4}$.}}\\
\vspace{0.2 cm}
\small{${}^1$ Max-Planck-Institut f\"ur Plasmaphysik, 85748 Garching, Germany\\
${}^2$ Max-Planck-Institut f\"ur Plasmaphysik, 17491 Greifswald, Germany\\
${}^3$ ENEA C. R. Frascati, Via E. Fermi 45, CP 65-00044 Frascati, Italy\\
${}^4$ Institute for Fusion Theory and Simulation, Zhejiang Univ., Hangzhou, People's Rep. of China\\
contact of main author: \url{www2.ipp.mpg.de/~biancala}
}
\end{center}

\begin{abstract}
The confinement of energetic particles (EP) is crucial for an efficient heating of tokamak plasmas. Plasma instabilities such as Alfv\'en Eigenmodes (AE) can redistribute the EP population making the plasma heating less effective, and leading to additional loads on the walls. The nonlinear dynamics of toroidicity induced AE (TAE) is investigated by means of the global gyrokinetic particle-in-cell code ORB5, within the NEMORB project.
The nonperturbative nonlinear interplay of TAEs and EP due to the wave-particle nonlinearity is studied. In particular, we focus on the nonlinear modification of the frequency, growth rate and radial structure of the TAE, depending on the evolution of the EP distribution in phase space. For the ITPA benchmark case, we find that the frequency increases when the growth rate decreases, and the mode shrinks radially. This nonlinear evolution is found to be correctly reproduced by means of a quasilinear model, namely a model where the linear effects of the nonlinearly modified EP distribution function are retained.
\end{abstract}

\section{Introduction.}
\label{sec:intro}

One of the main heating mechanisms for tokamaks plasmas is the injection of beams of energetic particles (EPs), whose task is the thermalization and transfer of energy to the bulk plasma. The process of thermalization is not immediate, and during the time that the EPs circulate inside the tokamak, they can drive plasma waves unstable via resonant interaction. Shear Alfv\'en waves (SAWs) are among the most unstable, because the characteristic Alfv\'en velocity $v_A=B/\sqrt{4\pi\varrho}$ ($B$ is the equilibrium magnetic field and $\varrho$ the mass density of the plasma) is comparable with that of the EPs. In addition, SAW group velocity is directed along the magnetic field line and, therefore, EPs can stay in resonance and effectively exchange energy with the wave~\cite{Zonca06,Chen16}.

SAW in a nonuniform equilibrium experience continuum damping~\cite{Hasegawa74,Chen74}, because of the formation of singular structures where the SAW continuum is resonantly excited. Two types of shear Alfv\'en modes exist in tokamak plasmas: energetic-particle continuum modes (EPM)~\cite{Chen94,Chen07}, with frequency determined by the EP characteristic frequencies, and discrete Alfv\'en eigenmodes (AE), with frequencies inside SAW continuum gaps~\cite{Cheng85}. EPMs can become unstable if the drive exceeds a threshold determined by the continuum damping absorption; AEs, on the other hand, are practically unaffected by continuum damping~\cite{Zonca06,Chen16,Hasegawa74,Chen74}, and therefore are generally more unstable. A unified approach for weakly and strongly driven AEs and EPMs has been recently derived, based on a generalized fishbone-like dispersion relation (GFLDR), which helps extracting the underlying physics of the numerical simulations~\cite{Zonca14b}.

The class of AEs of interest in this paper is named toroidicity-induced AEs (TAEs), whose frequency lies inside the continuum gap formed by the toroidal curvature~\cite{Cheng85,Fu89,Zonca92}.
The linear and nonlinear dynamics of TAEs has been the task of several numerical benchmarks, like the  International Tokamak Physics Activity (ITPA), comparing the results of several gyrokinetic (GK), gyrofluid, and hybrid GK-MHD codes~\cite{Koenies12}. A significant progress in the understanding of the wave-particle nonlinear interaction of TAEs has been recently achieved by means of hamiltonian-mapping techniques~\cite{Briguglio14}.

The numerical tool adopted for the studies presented here is the nonlinear gyrokinetic particle-in-cell (PIC) code ORB5~\cite{Jolliet07}, which now includes all extensions (including the electromagnetic and multi-species extensions) of the NEMORB project~\cite{Bottino11,Bottino15JPP,Tronko16,Biancalani16CPC}. The GK model contains the treatment of the resonances of ions and electrons, which is neglected by fluid models. The Lagrangian formulation that is used, is based on the GK Vlasov-Maxwell equations of Sugama, Brizard and Hahm~\cite{Sugama00,Brizard07}. The conservation theorems for the energy and momentum in GK framework~\cite{Scott10} are automatically satisfied due to the method of derivation of the GK Vlasov-Maxwell equations from a discretized Lagrangian~\cite{Bottino15JPP}.
As a consequence, this model can be adopted in principle for rigorous nonlinear electromagnetic simulations of global instabilities in the presence of EP and turbulence, where all nonlinearities are treated on the same footing in a self-consistent way.

The linear dynamics of the ITPA-TAE case has been investigated with ORB5 and described in Ref.~\cite{Biancalani16PoP} (like also done for example with the linear GK code GYGLES in Ref.~\cite{Mishchenko09}). In particular, the radial structure of the mode has been found to depend not only on the interaction with the EP population, but on the position of the frequency with respect to the gap in the continuum. In the linear regime, the dependence of the AE mode structure on the EP distribution function had also been previously studied analytically~\cite{Ma15} and by means of gyrofluid and GK simulations, with a comparison with experiments~\cite{Tobias11,Spong11}, and discussed in the framework of the GFLDR theory~\cite{Zonca14b}. In this paper, we extend the linear work done with ORB5 in Ref.~\cite{Biancalani16PoP} to the investigation of the nonlinear wave-particle interaction (see also Ref.~\cite{Cole16}, where a comparison on the TAE nonlinear saturation level with different models is shown). In particular, 
the nonlinear modification of the frequency, growth rate and radial structure is 
investigated here. 

The paper is organized as follows. Section~\ref{sec:model} is devoted to the description of the model. The equilibrium is given in Section~\ref{sec:equil}. The linear dynamics as depending on the position of the frequency in the continuum spectrum is recalled in Section~\ref{sec:Lin}. The nonlinear modification of the frequency, growth rate and radial mode structure is described in Section~\ref{sec:NL-QL}, where a quasilinear analysis, helping understanding the results of the nonlinear investigation, is also provided. Finally, a summary of the results is given in Section~\ref{sec:conclusions}.

\section{The model}
\label{sec:model}

The gyrokinetic Lagrangian (see Ref.~\cite{Bottino15JPP} and references therein) is the starting point for the derivation of the model equations of ORB5:
\begin{equation}
 \mathscr{L} = \Sigma_{\rm{sp}}\int \d V \d W \Big[ \Big( \Big( \frac{e}{c}\vecb{A}+p_\|\vecb{b} \Big)      \cdot\dot{\vecb{R}}+\frac{m  c}{e}\mu \dot{\theta} \Big) f -\big( \mathscr{H}_0 + \mathscr{H}_1 \big) f -  \mathscr{H}_2 f_M \Big] - \int \d V \frac{|\nabla_\perp A_\parallel|^2}{8\pi} \label{eq:Lagrangian}
\end{equation}
where the Hamiltonian is divided into unperturbed, linear, and nonlinear part, $\mathscr{H} = \mathscr{H}_0 +\mathscr{H}_1 + \mathscr{H}_2 $, with:
\begin{eqnarray}
 \mathscr{H}_0 & = & \frac{p_\|^2}{2m} + \mu B \\
 \mathscr{H}_1 & = & e \Big(  J_0\Phi - \frac{p_\|}{mc} J_0 A_\parallel \Big)  \\
 \mathscr{H}_2 & = & \frac{e^2}{2mc^2} (J_0 A_\parallel)^2  - \frac{mc^2}{2B^2}|\nabla_\perp\phi|^2
\end{eqnarray}
Here $f$ and $f_M$ are the total and equilibrium (i.e. independent of time) distribution functions, the integrals are over the phase space volume, with $\d V$ being the real-space infinitesimal and $\d W = (2\pi/m^2) B_\|^* \d p_\| \d \mu$ the velocity-space infinitesimal.
The phase-space coordinates are $\vecb{Z}=(\vecb{R},p_\|,\mu)$, i.e. respectively the gyrocenter position, canonical parallel momentum $p_\| = m U + (e/c) J_0 A_\|$ and magnetic momentum $\mu = m v_\perp^2 / (2B)$. The Jacobian is given by the parallel component of $\vecb{B}^*= \vecb{B} + (c/e) p_\| \vecb{\nabla}\times \vecb{b}$, where $\vecb{B}$ and $\vecb{b}$ are the equilibrium magnetic field and magnetic unit vector.
The time-dependent fields are named $\phi$ and $A_\|$, and they are respectively the perturbed scalar potential and the parallel component of the perturbed vector potential.
In our notation, on the other hand, $\vecb{A}$ is the equilibrium vector potential.  The summation is over all species present in the plasma, and the gyroaverage operator is labeled here by $J_0$ (with $J_0=1$ for electrons). The gyroaverage operator reduces to the Bessel function if we transform into Fourier space.

The model equations of ORB5 are the gyrocenter trajectories, and the two equations for the fields~\cite{Bottino15JPP}. They are derived by imposing the minimal action principle with respect to the phase-space coordinates $(\vecb{R},p_\|,\mu)$, and to $\phi$ and $A_\|$. The gyrocenter trajectories are:
\begin{eqnarray}
\dot{\vecb  R}&=&\frac{1}{m}\left(p_\|-\frac{e}{c}J_0A_\parallel\right)\frac{\vecb{B^*}}{B^*_\parallel} + \frac{c}{e B^*_\parallel} \vecb{b}\times \left[\mu
  \nabla B + e \nabla J_0  \big(\phi -  \frac{p_\|}{mc} A_\| \big) \right] \label{eq:trajectories_a} \\
\dot{p_\|}&=&-\frac{\vecb{B^*}}{B^*_\parallel}\cdot\left[\mu \nabla B + e
  \nabla J_0 \big(\phi -  \frac{p_\|}{mc} A_\| \big) \right] \label{eq:trajectories_b}
\end{eqnarray}
The GK Poisson equation is:
\begin{equation}
 - \vecb{\nabla} \cdot \frac{mc^2\int \d W f_M}{B^2} \nabla_\perp \phi=\Sigma_{\rm{sp}} \int \d W  e J_0 f  \label{eq:Poisson}
\end{equation}
The Amp\`ere equation is:
\begin{eqnarray}
\Sigma_{\rm{sp}} \int \d W \Big( \frac{ep_\|}{mc} f-\frac{e^2}{mc^2}A_\parallel f_{M}
 \Big)  +  \frac{1}{4\pi}\nabla_\perp ^2 A_\parallel =0 \label{eq:Ampere}
\end{eqnarray}
Here the form with $J_0=1$ is given for simplicity, also in the view of the comparison with MHD codes. For more complicated models, see Ref.~\cite{Bottino15JPP,Tronko16}. In the simulations shown in this paper, a Fourier filter in mode numbers is applied, keeping only $n=6$ and $m=9,10,11,12$. Dirichlet boundary conditions are imposed to the potentials.

In PIC codes written in $p_\|$ formulation, a numerical problem called ``cancellation problem''~\cite{Mishchenko04,Hatzky07} arises in particular in the numerical resolution of the Amp\`ere's equation. This is due to fact that the statistical error affects only the term discretized with markers (first term in Eq.~\ref{eq:Ampere}), and the result is a numerical error which can be orders of magnitude higher than the desired signal. 
A solution to this ``cancellation problem'' comes with a control-variate technique, which splits the perturbed distribution function $\delta f$ in an adiabatic part $\delta f^{ad} = - (J_0 \phi - p_\| J_0 A_\|/mc) e f_M / k_B T$ and a nonadiabatic part (i.e. the remaining part). With this technique, the integral to be performed with the marker discretization becomes in fact much smaller, and therefore the resulting numerical noise is greatly mitigated~\cite{Mishchenko04,Hatzky07}.
This control-variate technique has been recently implemented in ORB5, making numerical simulations of electromagnetic instabilities possible~\cite{Bottino11,Biancalani16CPC}.

\section{Equilibrium}
\label{sec:equil}

The equilibrium of the International Tokamak Physics Activity (ITPA) TAE case~\cite{Koenies12} is considered in this paper, like in Ref.~\cite{Biancalani16PoP,Mishchenko09}. The major radius and minor radius are $R = 10$ m and $a = 1$ m. The toroidal magnetic field is $B_0=3$ T, and the safety factor is $q(r) = 1.71 + 0.16 (r/a)^2$. The bulk ion is chosen to be hydrogen. The ion and electron average densities are $n_i = n_e = 2\cdot 10^{19} $m$^{-3}$. In the case of a nonuniform profile of an additional species (for example of the EP), the bulk ion and electron profiles are corrected in order to satisfy quasi-neutrality, as described below. The bulk ion and electron temperature is $T_i = T_e = 1$ keV.
The corresponding bulk ion cyclotron frequency is $\Omega_i = 2.87 \cdot 10^8$ rad/s, the thermal ion velocity is $v_{th,i} = \sqrt{T_i/m_i} = 3.095\cdot 10^5$ m/s and the sound Larmor radius is $\rho_s = \sqrt{T_e/m_i}/\Omega_i = 1.078\cdot 10^{-3}$ m. The electron beta on axis is $\beta_e = 8 \pi n_e T_e/ B_0^2 = 8.955 \cdot 10^{-4}$ and the Alfv\'en velocity on axis is $v_A = 1.46\cdot 10^7$ m/s.

\begin{figure}[b!]
\begin{center}
\includegraphics[width=0.44\textwidth]{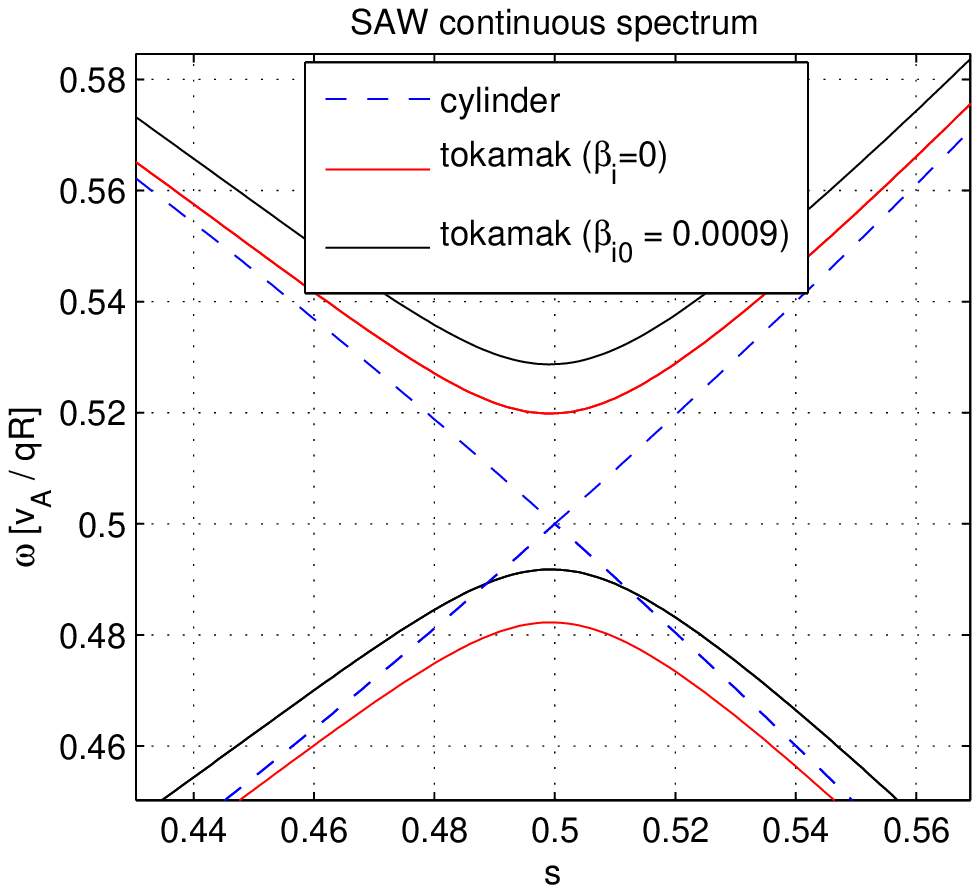}
\includegraphics[width=0.445\textwidth]{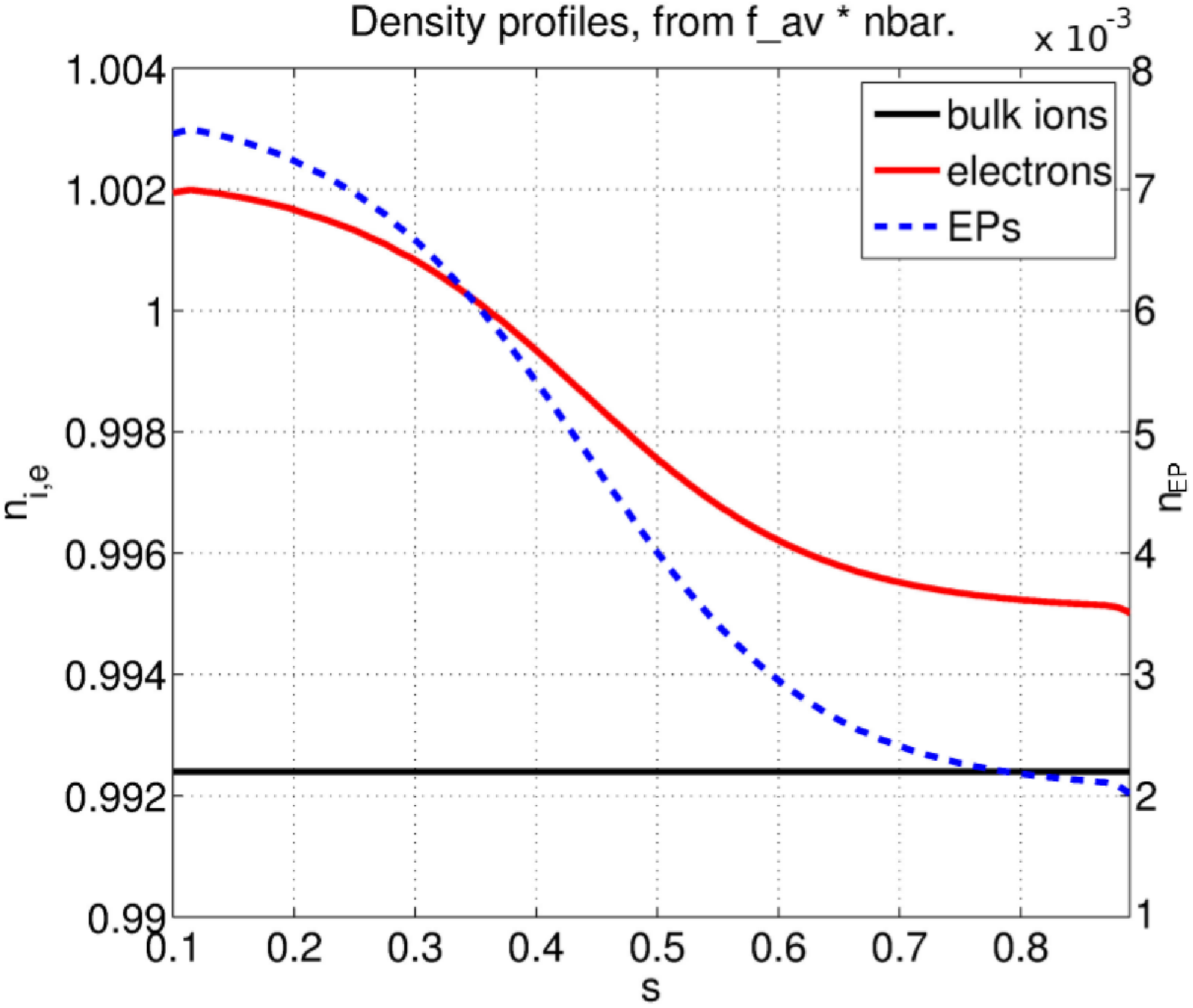}
\vskip -1em
\caption{On the left, zoom of the continuous spectrum obtained by neglecting toroidicity and compressibility (blue dashed lines), or by keeping toroidicity only (red line) or by keeping both toroidicity and compressibility (black line), with approximated formula of Ref.~\cite{Biancalani16PoP}. On the right, density profiles in units of $n_{e0}=2\cdot10^{19} m^{-3}$, for a case with $\langle n_{EP}\rangle / n_{e0}=0.004$.}\label{fig:LIN-continuum}
\end{center}
\end{figure}

Given these parameters, we can calculate the continuous spectrum near the toroidicity induced gap of the TAE with toroidal mode number $n=6$ (see Fig.~\ref{fig:LIN-continuum}). The X-point where the two cylinder continuum branches with $m=10,11$ cross (neglecting toroidicity and compressibility) is located at $\omega_{CXP} = 0.5 \,v_A / q R = 4.17\cdot 10^5$ rad/s = $1.454\cdot 10^{-3} \Omega_i$ (where q=1.75). The continuum accumulation points can be calculated by using the approximated formula given in Ref.~\cite{Biancalani16PoP} (derived after Ref.~\cite{Cheng85,Fu89}), which is valid for small values of the inverse aspect ratio, and slightly underestimates the gap width.
We obtain that the lower continuum accumulation point (LCAP) calculated with compressibility effects, is located at $\omega_{LCAP} \simeq 4.1\cdot 10^5$ rad/s $=0.492 \, v_A/qR$ and the upper continuum accumulation point (UCAP) at $\omega_{UCAP} \simeq 4.4\cdot 10^5$ rad/s $= 0.53 \, v_A/qR$.
The upshift of the frequency due to compressibility has been estimated by using the beta-induced Alfv\'en eigenmode CAP calculated in Ref.~\cite{Zonca96}: $\omega(T) = (\omega_{T=0}^2+\omega_{BAE-CAP}^2)^{1/2}$, with $\omega_{BAE-CAP} = 0.8 \cdot 10^5$ rad/s $= 0.1 \, v_A/qR$. In the proximity of the TAE frequency, this gives an upshift of about $0.9\cdot 10^4$ rad/s $\simeq 0.01 \, v_A/qR$ (neglected in the calculation of the CAPs in Ref.~\cite{Mishchenko09}).

A distribution function Maxwellian in velocity-space is considered for a population of deuterium, playing the role of the EP species. The EP averaged concentration of a reference case is $\langle n_{EP} \rangle /n_e = 0.004$ with radial profile given by:
\begin{equation}\label{eq:TAE-n_EP}
n_{EP}(s)/n_{EP}(s_0) = \exp [-\Delta \, \kappa_n \tanh ((s-s_0)/\Delta)] 
\end{equation}
with $s_0=0.5$, $\Delta=0.2$, and $\kappa_n = 3.333$. The radial coordinate here is $s=\sqrt{\psi/\psi_{edge}}\simeq r/a$, with $\psi$ being the poloidal magnetic equilibrium flux. When the EP ion species is added, the total number of electrons in the volume is kept the same, and the bulk ion average concentration is diminished in order for the total number of ions (bulk plus energetic) to equal the total number of electrons.
The bulk ions profile is kept flat, as in the absence of EP. The electron profile is shaped (see Fig.~\ref{fig:LIN-continuum}) in the simulations used in this paper, in order to satisfy $n_e(r) = \sum_i Z_i  n_i(r)$ (whereas the electron profile was kept flat in Ref.~\cite{Biancalani16PoP}). The reference EP temperature in this paper is $T_{EP}=400$ keV (whereas most of the TAE simulations in Ref.~\cite{Biancalani16PoP} used $T_{EP}=500$ keV).
No finite-Larmor-radius (FLR) effects are retained in the simulations shown in this paper, consistently with Ref.~\cite{Biancalani16PoP}. This means that the Bessel functions are calculated for vanishing argument. The investigation of the FLR effects will be done in a separate paper. Simulations with $m_i/m_e$ = 200 are considered in this paper. Convergence tests are given in Appendix~\ref{sec:appendix_conv_tests}.

\section{Linear dynamics}
\label{sec:Lin}

In this section, we investigate numerically the dynamics of linear simulations, namely simulations where all three species follow unperturbed trajectories.

When an initial perturbation is let evolve in time in the absence of EP, a TAE is observed as a discrete global mode with frequency lying within the gap. The measured frequency is: $\omega= 0.493 \, v_A/qR$.

\subsection{Dependence on the EP concentration}
\label{sec:Lin-vs-nEP}

\begin{figure}[b!]
\begin{center}
\includegraphics[width=0.43\textwidth]{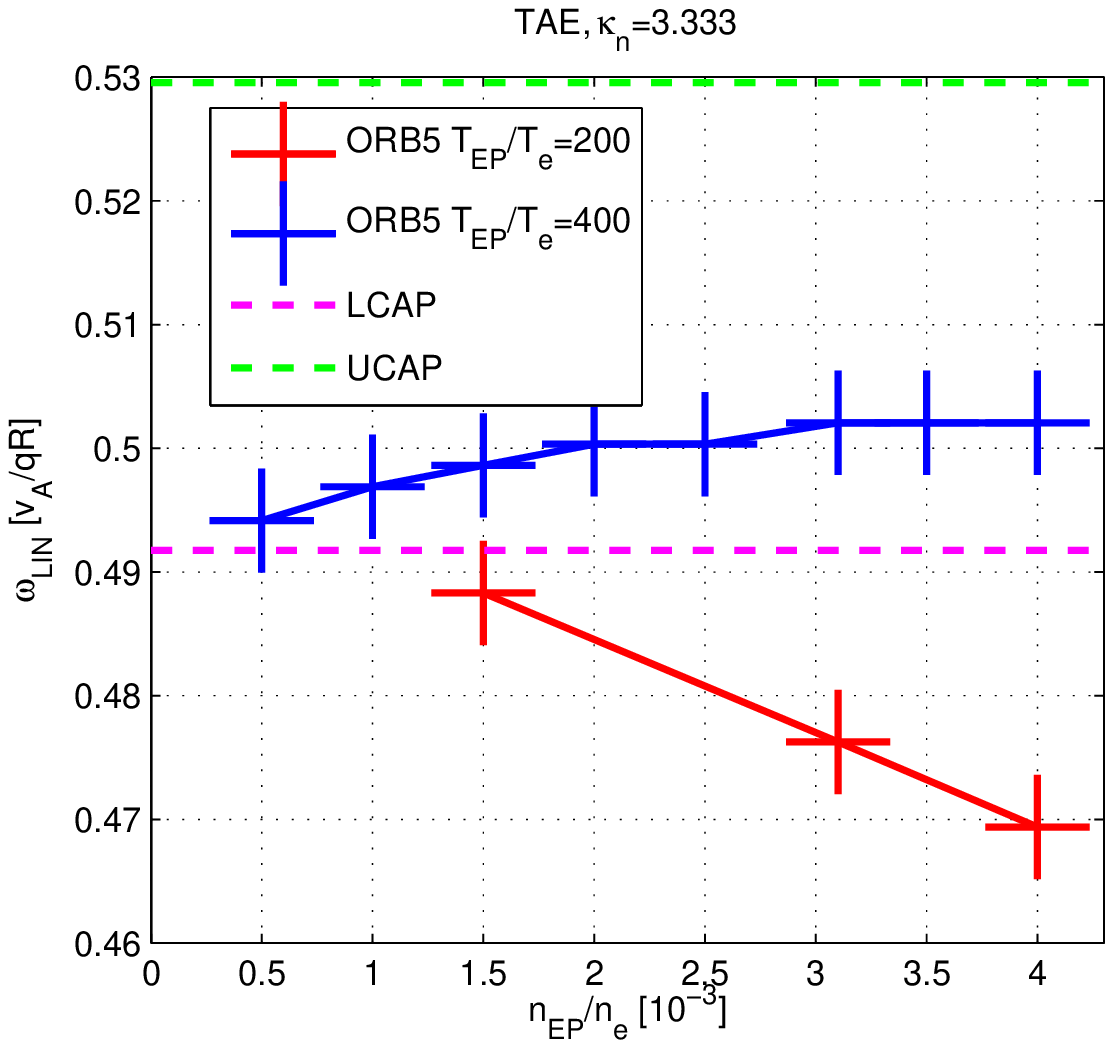}
\includegraphics[width=0.44\textwidth]{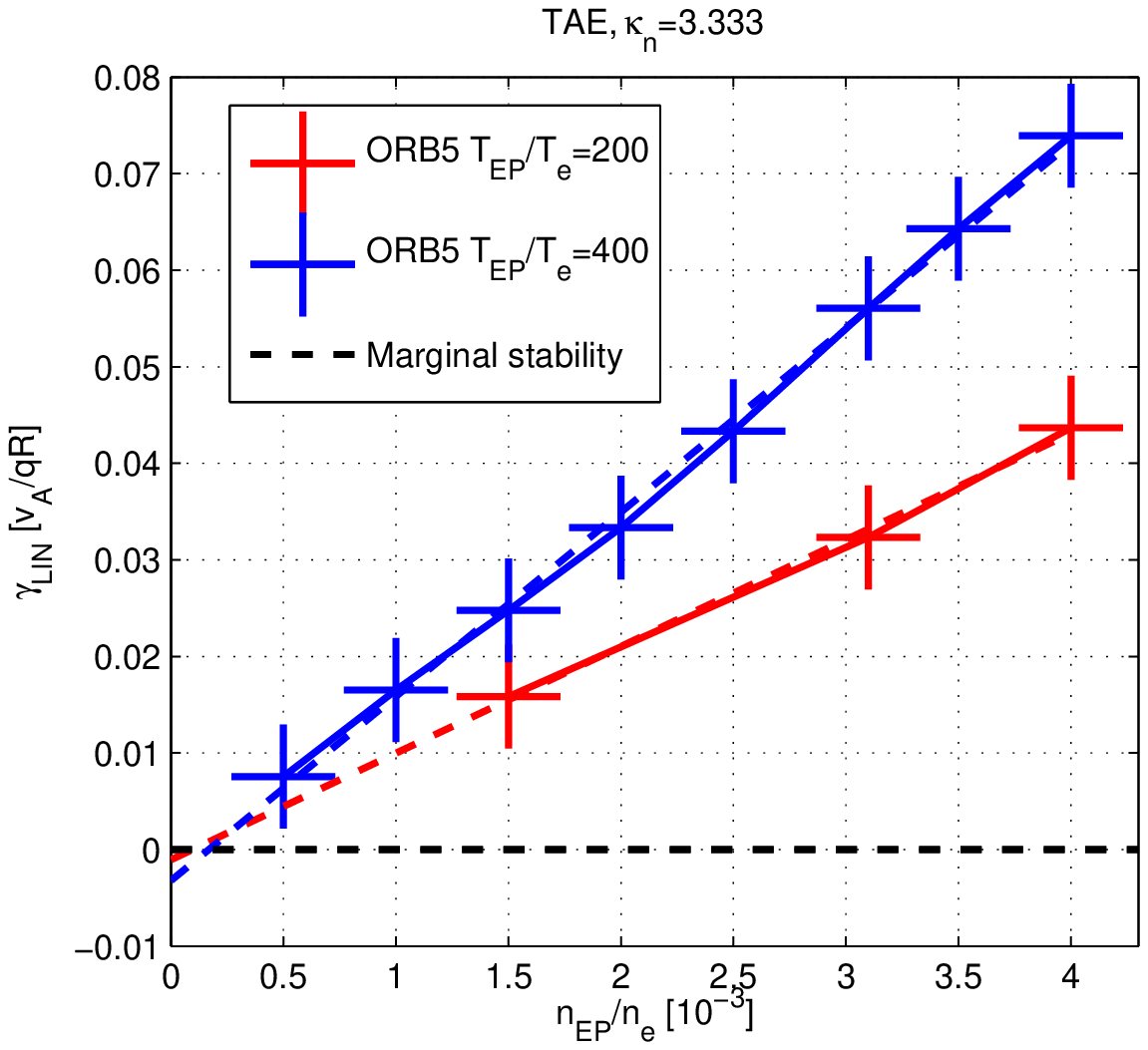}
\vskip -1.em
\caption{Frequency (left) and growth rate (right) dependence on EP average density, for two different values of $T_{EP}$. The frequencies of the lower and upper CAPs with compressibility effects are also plotted as magenta and green dashed lines.}\label{fig:LIN-omegagamma_nEP}
\end{center}
\end{figure}

Two temperatures of the EP populations have been considered: $T_{EP}/T_e=200$ and $T_{EP}/T_e=400$. In the former case, with increasing EP concentration, we observe a decreasing value of the frequency, going below the LCAP. On the contrary, in the latter case, the frequency slowly increases with EP concentration. Both cases show a clearly linearly increasing value of growth rate with the EP concentration. The damping is measured as $\gamma(n_{EP}\rightarrow 0)\sim - 10^{-3}  \, v_A/qR$, and is consistent with the analytical estimations of the electron Landau damping  given in Ref.~\cite{Fu89,Zonca92}.

\begin{figure}[t!]
\begin{center}
\includegraphics[width=0.47\textwidth]{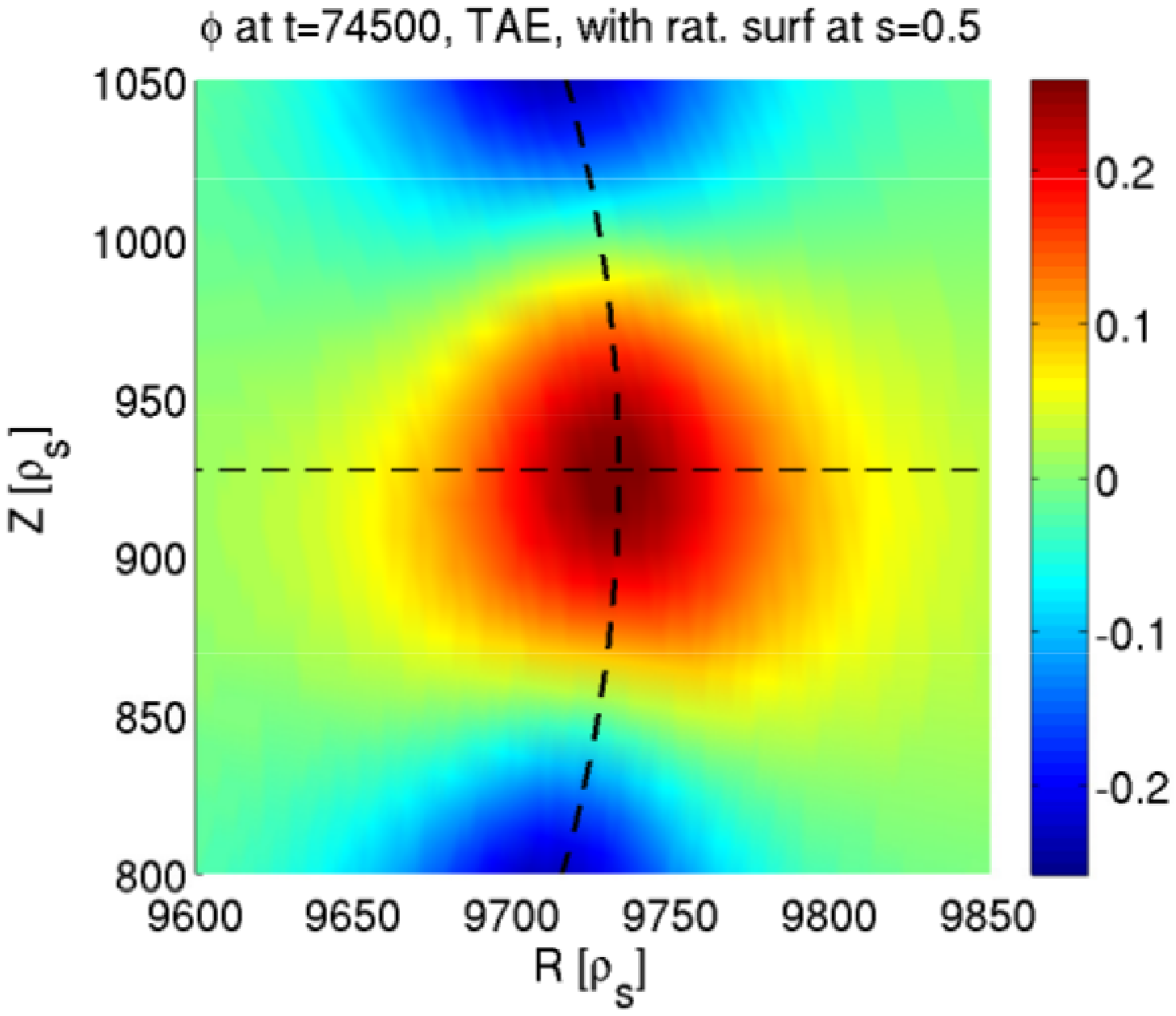}
\includegraphics[width=0.47\textwidth]{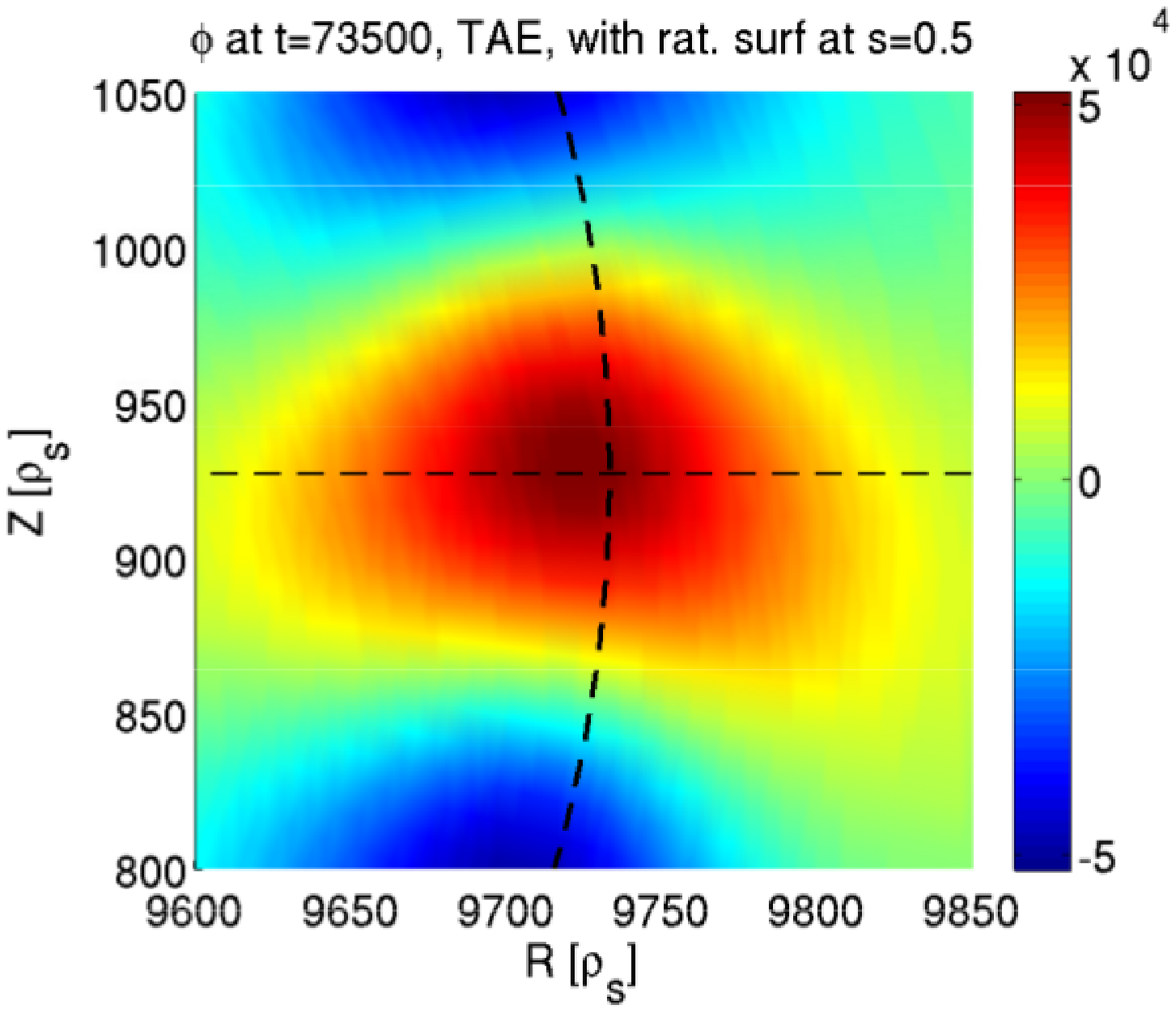}
\vskip -1.5em
\caption{Structure in the poloidal plane, for $n_{EP}/n_e = 0.001$ (left) and $n_{EP}/n_e = 0.004$ (right). Here $T_{EP}/T_e = 400$, and $\kappa_n = 3.333$.}\label{fig:LIN-structure_nEP}
\end{center}
\end{figure}

When considering modes with frequency inside the toroidicity induced gap (like for example for $T_{EP}/T_e=400$), the structure of the TAE in the poloidal plane does not show any ``boomerang'' (a.k.a. ``croissant'') shape, because of the weak interaction with the continuum, consistently with what shown in Ref.~\cite{Biancalani16PoP}. When increasing the EP concentration, we observe that the radial width increases (see Fig.~\ref{fig:LIN-structure_nEP}) as the region where the drive overcomes the damping increases. Scalings similar to those obtained in Ref.~\cite{Biancalani16PoP}, namely without imposing the quasineutrality at the initial time, are shown in Appendix~\ref{sec:appendix_flat}.

\subsection{Dependence on the EP temperature}
\label{sec:Lin-vs-TEP}

\begin{figure}[b!]
\begin{center}
\includegraphics[width=0.45\textwidth]{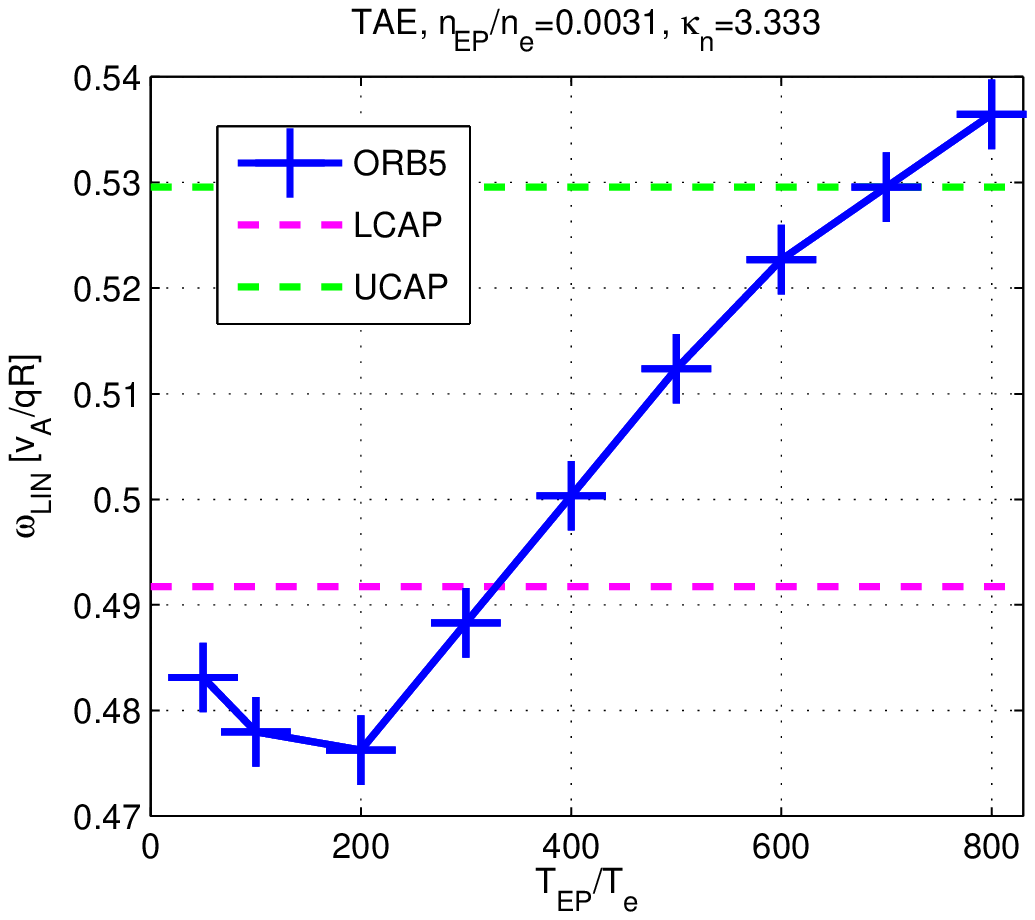}
\includegraphics[width=0.45\textwidth]{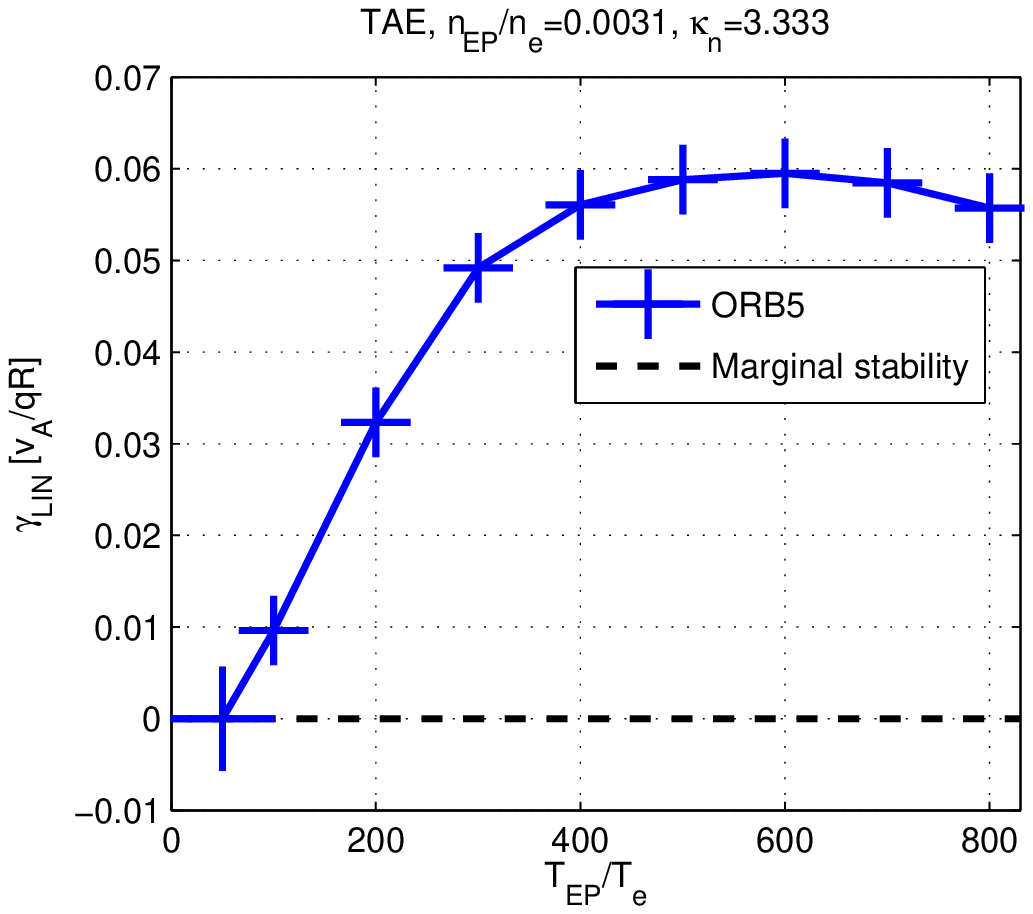}
\vskip -1em
\caption{Frequency (left) and growth rate (right) dependence on EP temperature. The frequencies of the lower and upper CAPs with compressibility effects are also plotted as magenta and green dashed lines. The analytical predictions for the growth rate are also shown.}\label{fig:LIN-omegagamma_TEP}
\end{center}
\end{figure}

By considering $n_{EP}/n_e = 0.0031$ and increasing the EP temperature, we note that at low temperatures the frequency decreases, reaches the minimum of $\omega\simeq 0.477 \, v_A/qR$ at $T_{EP}/T_e>200$ (consistently with Ref.~\cite{Mishchenko09}), then grows linearly and enters the continuum gap, and the growth rate increases overcoming the instability threshold at $T_{EP}/T_e \simeq 50$, and then reaches a maximum value of $\gamma\simeq 0.06\, v_A/qR$ at $T_{EP}/T_e \simeq 600$  (consistently with Ref.~\cite{Mishchenko09}, see Fig.~\ref{fig:LIN-omegagamma_TEP}).

\begin{figure}[t!]
\begin{center}
\includegraphics[width=0.49\textwidth]{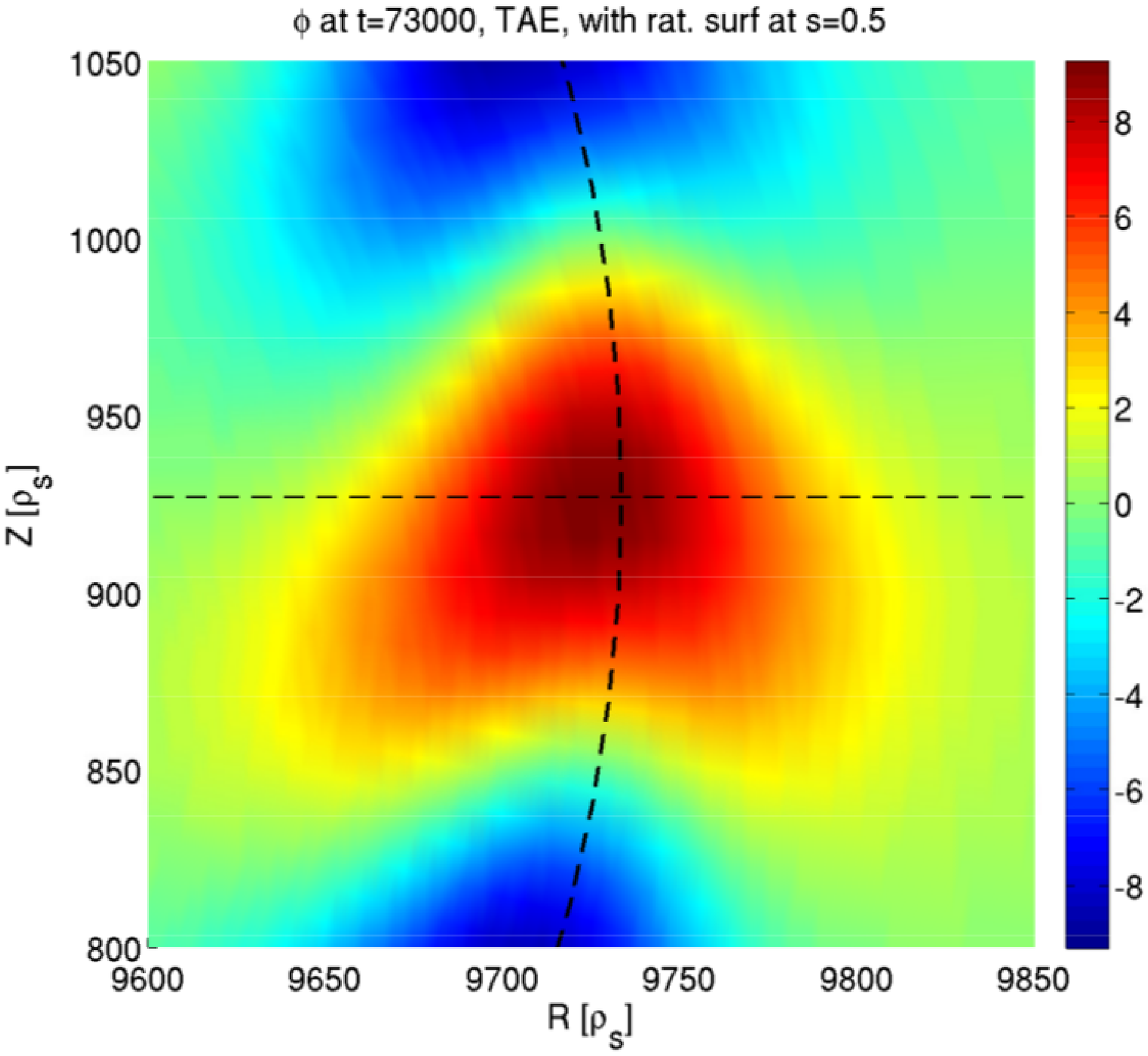}
\includegraphics[width=0.49\textwidth]{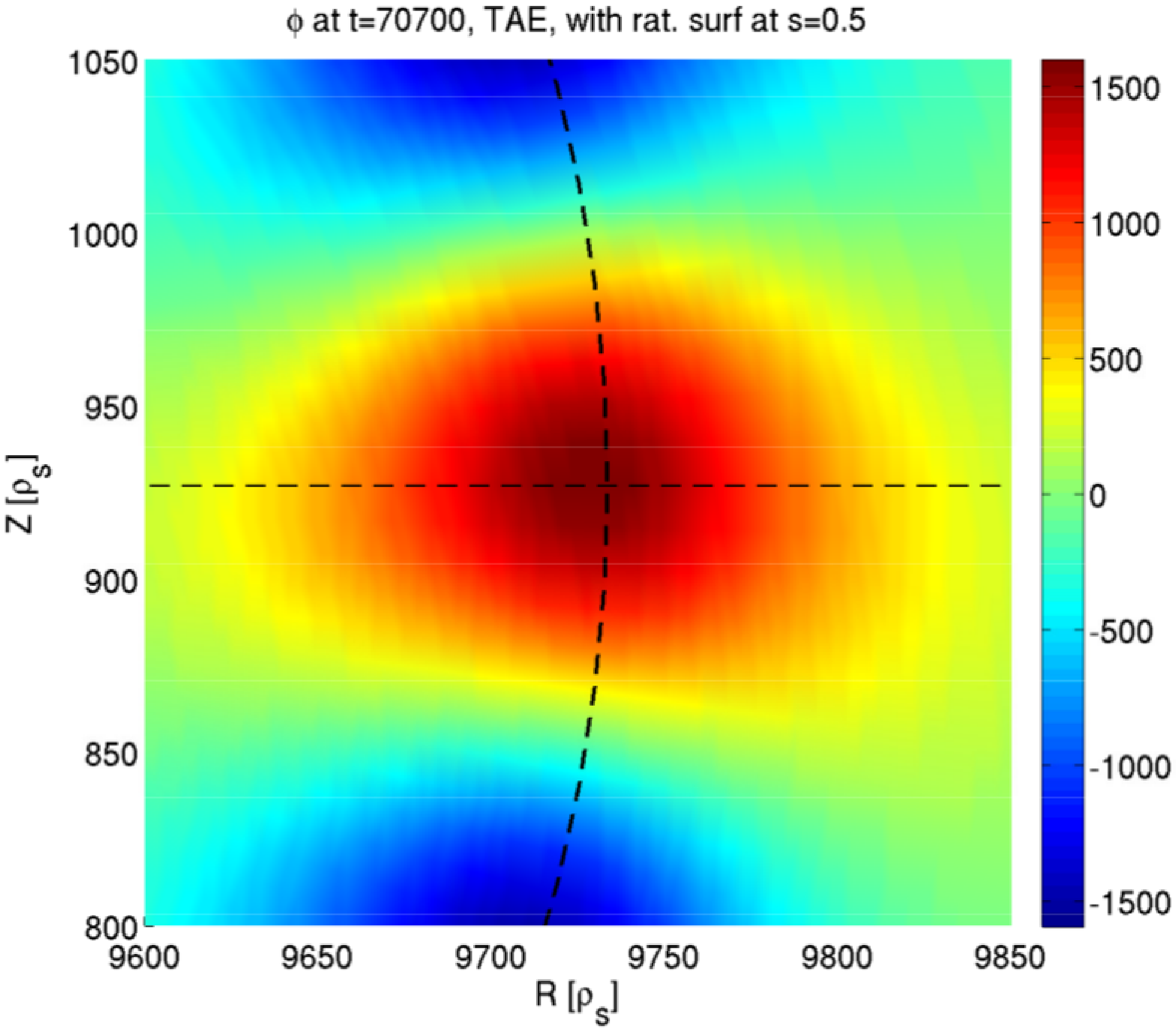}
\vskip -1em
\caption{Structure in the poloidal plane, for $T_{EP}/T_e = 200$ (left) and $T_{EP}/T_e = 600$ (right). Here $n_{EP}/n_e = 0.0031$, and $\kappa_n = 3.333$.}\label{fig:LIN-structure_TEP}
\end{center}
\end{figure}

At low EP temperatures, the structure in poloidal plane has a characteristic boomerang shape which reflects the fact that the frequency is well below the LCAP (see Ref.~\cite{Biancalani16PoP}). On the other hand, for higher temperatures, the freq. is in the gap and the boomerang shape disappears (see Fig.~\ref{fig:LIN-structure_TEP}).

\subsection{Dependence on the EP density gradient}
\label{sec:Lin-vs-gradnEP}

\begin{figure}[b!]
\begin{center}
\includegraphics[width=0.46\textwidth]{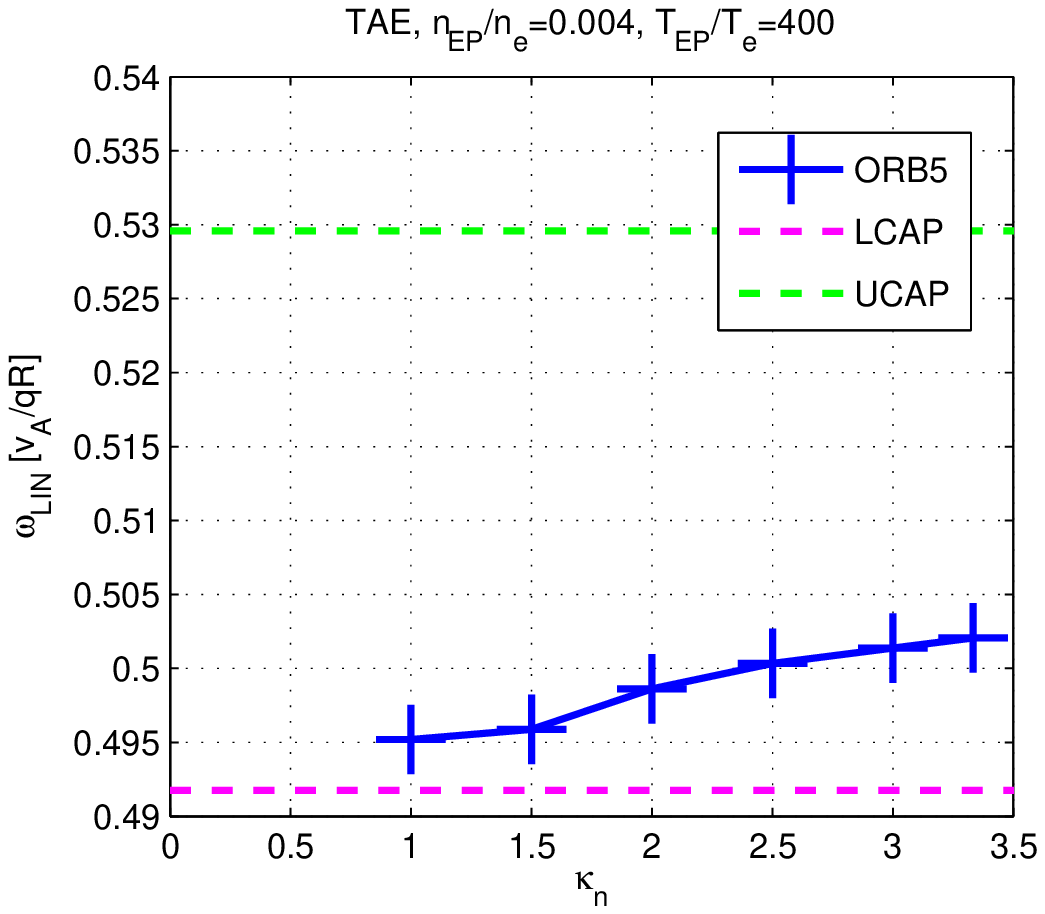}
\includegraphics[width=0.475\textwidth]{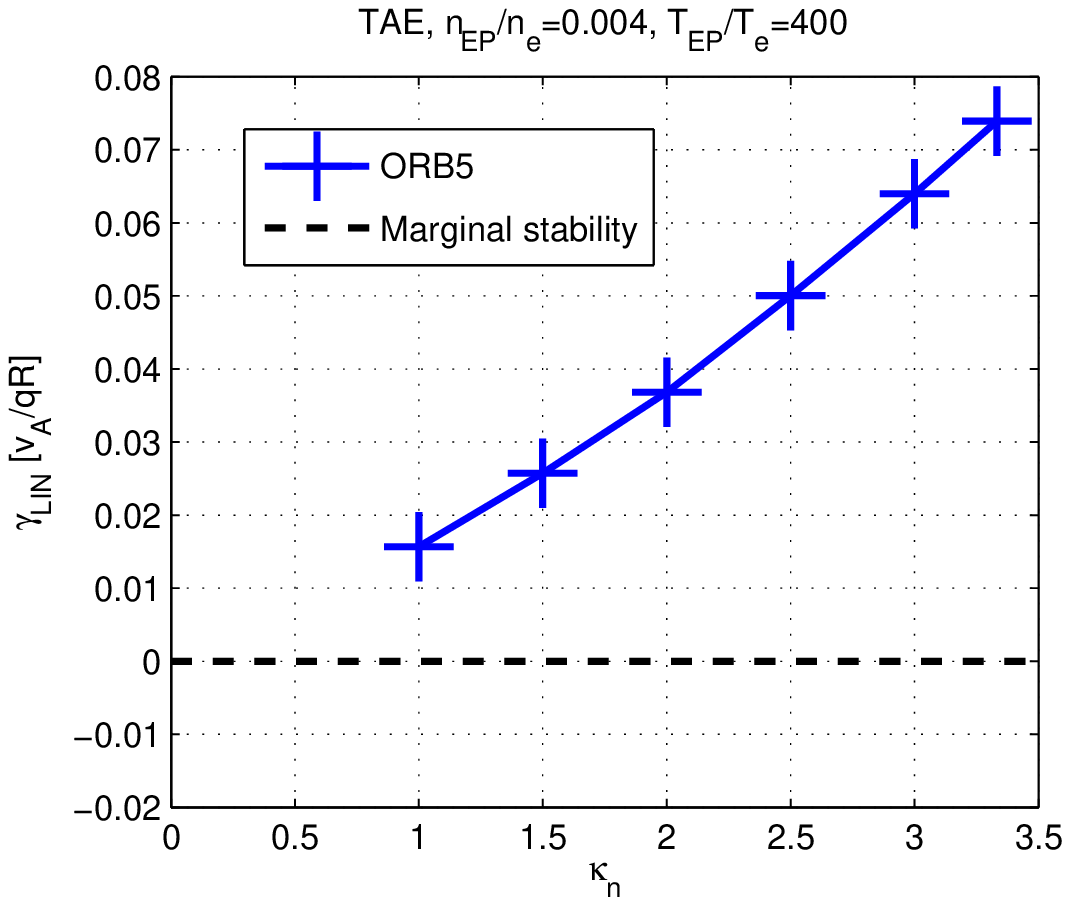}
\vskip -1em
\caption{Frequency (left) and growth rate (right) dependence on EP density gradient, for simulations with $T_{EP}/T_e = 400$. The frequencies of the lower and upper CAPs with compressibility effects are also plotted as magenta and green dashed lines.}\label{fig:LIN-omegagamma_gradnEP}
\end{center}
\end{figure}

With increasing EP density gradient, the frequency and the growth rate are observed to increase linearly (see Fig.~\ref{fig:LIN-omegagamma_gradnEP}).
The linear dependence of the frequency and growth rate on the EP density gradient is very similar to the dependence on the EP density (consistently with Ref.~\cite{Fu89}).

\begin{figure}[h!]
\begin{center}
\includegraphics[width=0.45\textwidth]{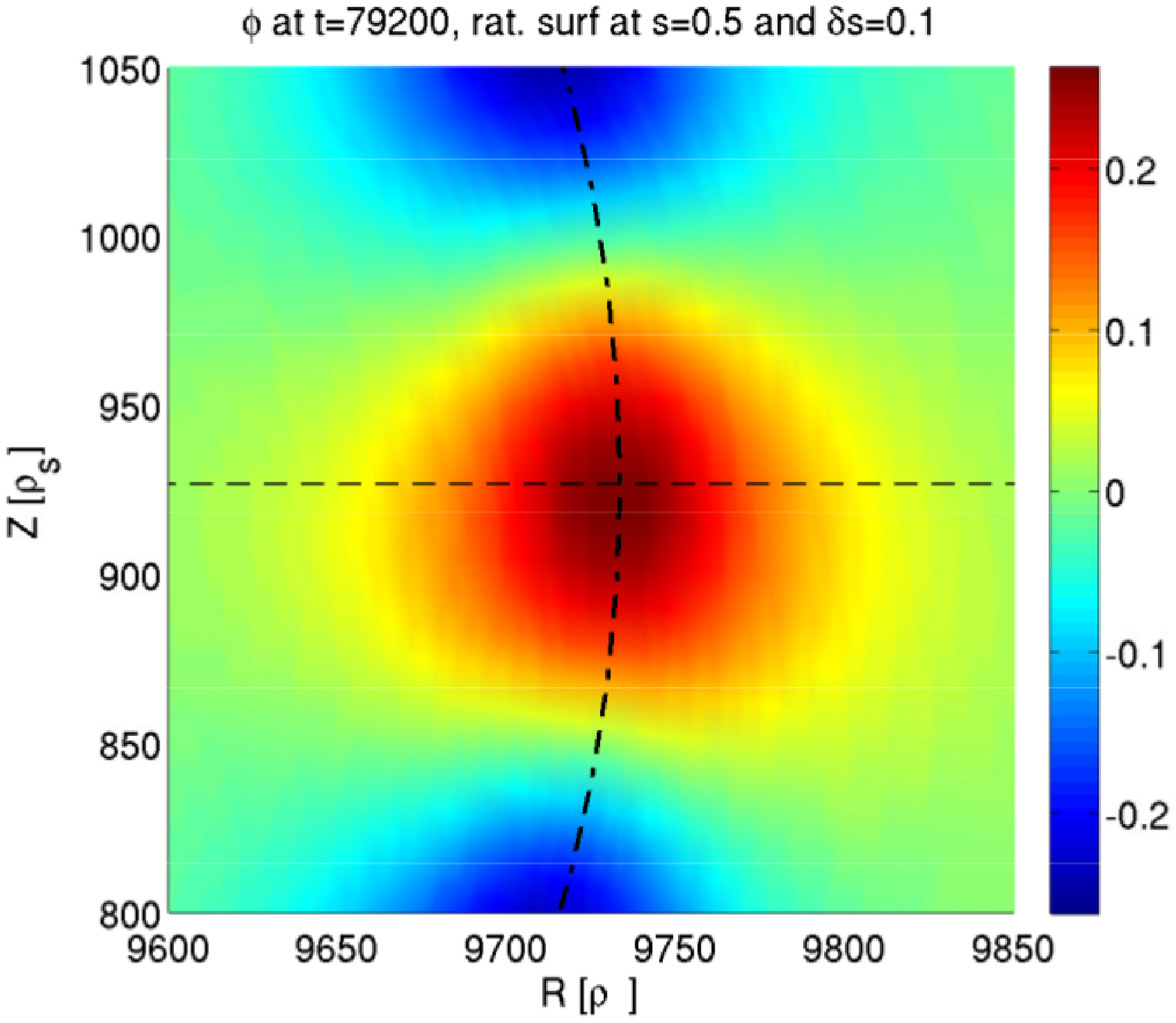}
\includegraphics[width=0.45\textwidth]{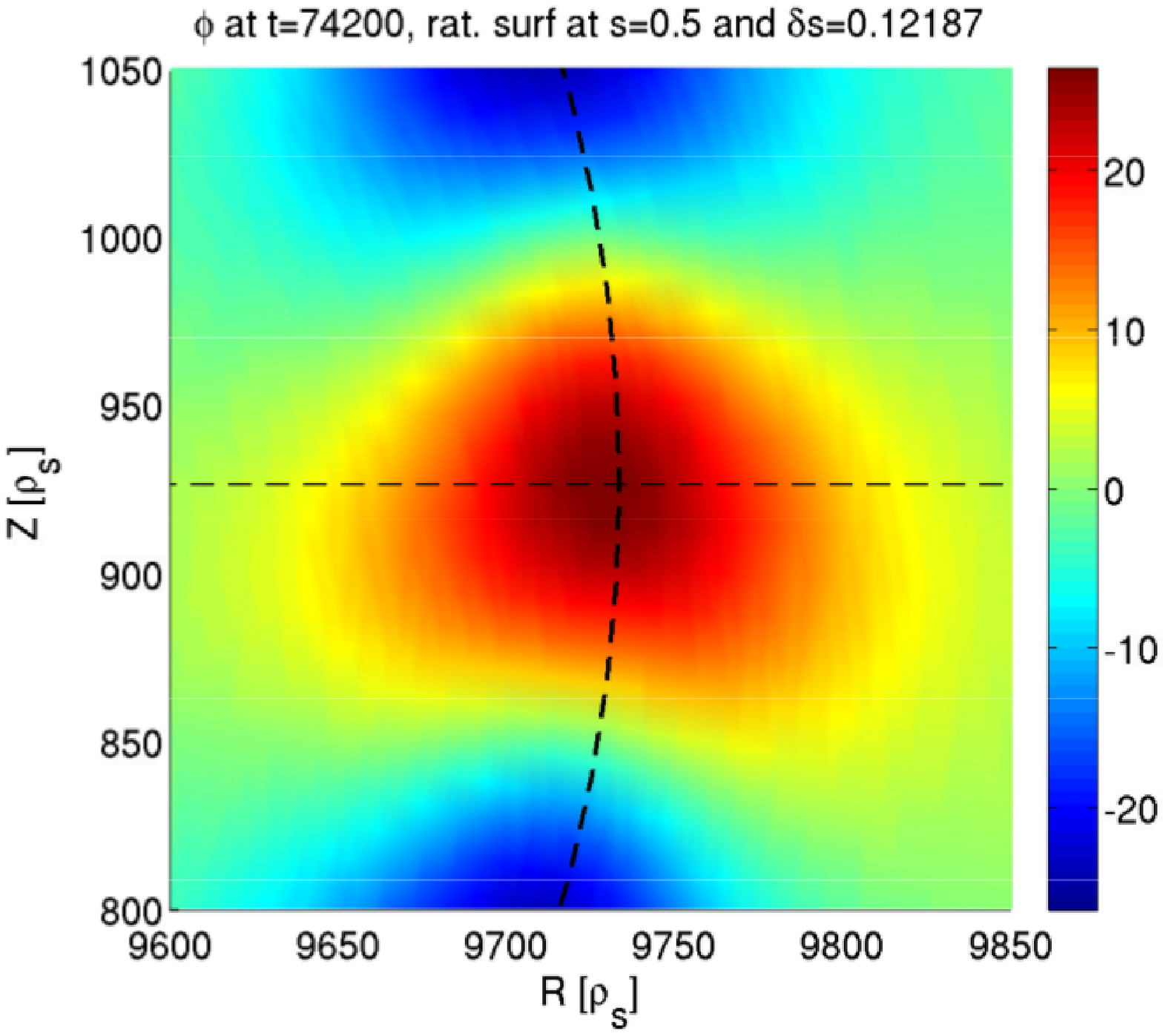}
\vskip -1em
\caption{Structure in the poloidal plane, for $\kappa_n = 1.0$ (left) and $\kappa_n = 2.0$ (right). Here $n_{EP}/n_e = 0.004$, $T_{EP}/T_e = 400$.}\label{fig:LIN-structure_gradnEP}
\end{center}
\end{figure}

Regarding the mode structure, when keeping the local concentration of EP constant and increasing the EP density gradient, we observe a similar behavior as when increasing the EP density, namely the radial width increases (see Fig.~\ref{fig:LIN-structure_gradnEP}).

\section{Nonlinear dynamics and quasilinear analysis}
\label{sec:NL-QL}

In this section, we consider a typical simulation with nonlinear wave-particle interaction. In a PIC code, this means that the bulk ions and electrons follow unperturbed trajectories, whereas the EP follow perturbed trajectories. We focus here on the nonlinear frequency and structure modification. The investigation of the dependence of the saturation levels over $n_{EP}$ is discussed in Ref.~\cite{Cole16}, where the results of the GK PIC codes ORB5 and EUTERPE are shown.

\begin{figure}[b!]
\begin{center}
\includegraphics[width=0.45\textwidth]{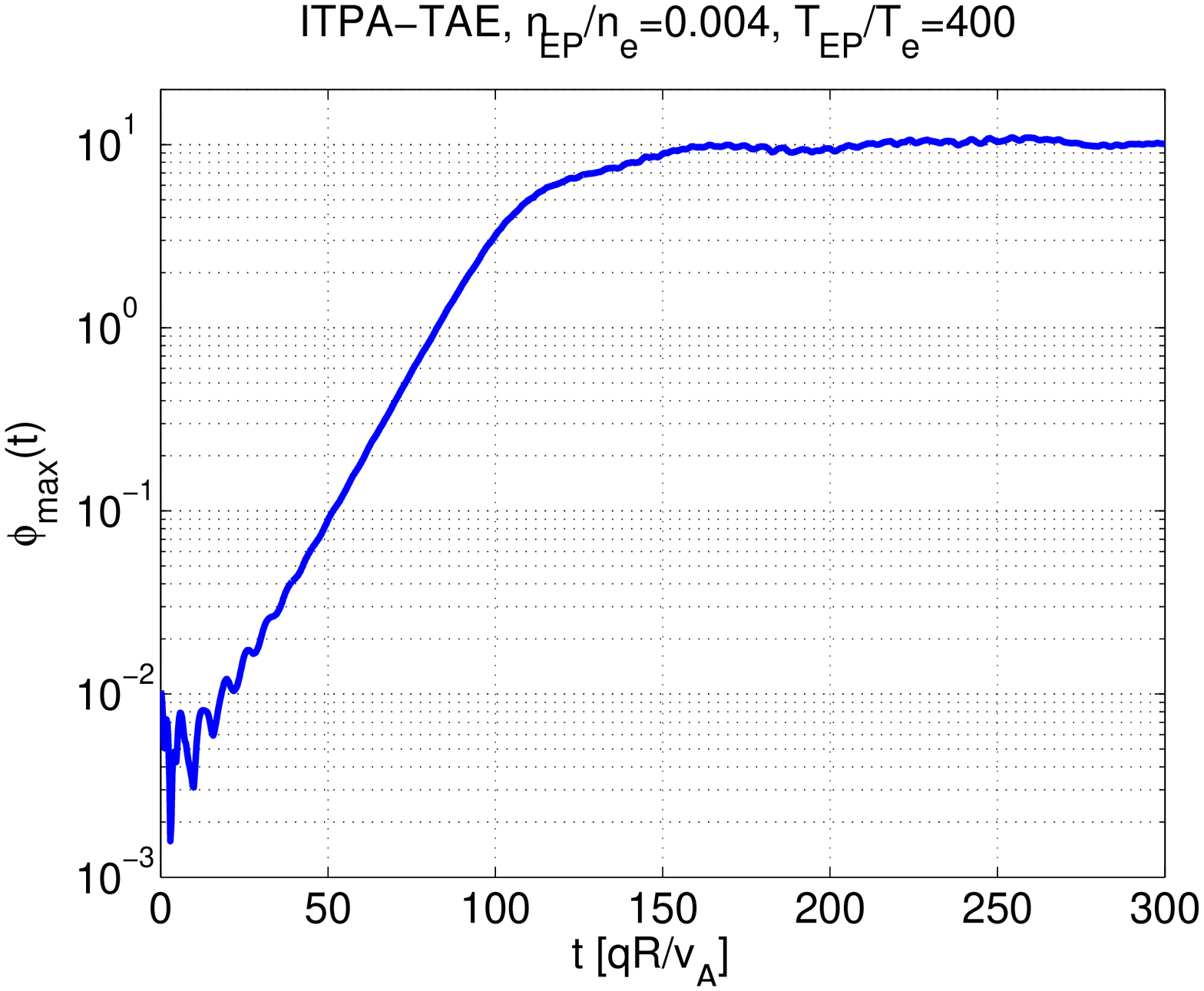}
\raisebox{-0.06\height}{\includegraphics[width=0.42\textwidth]{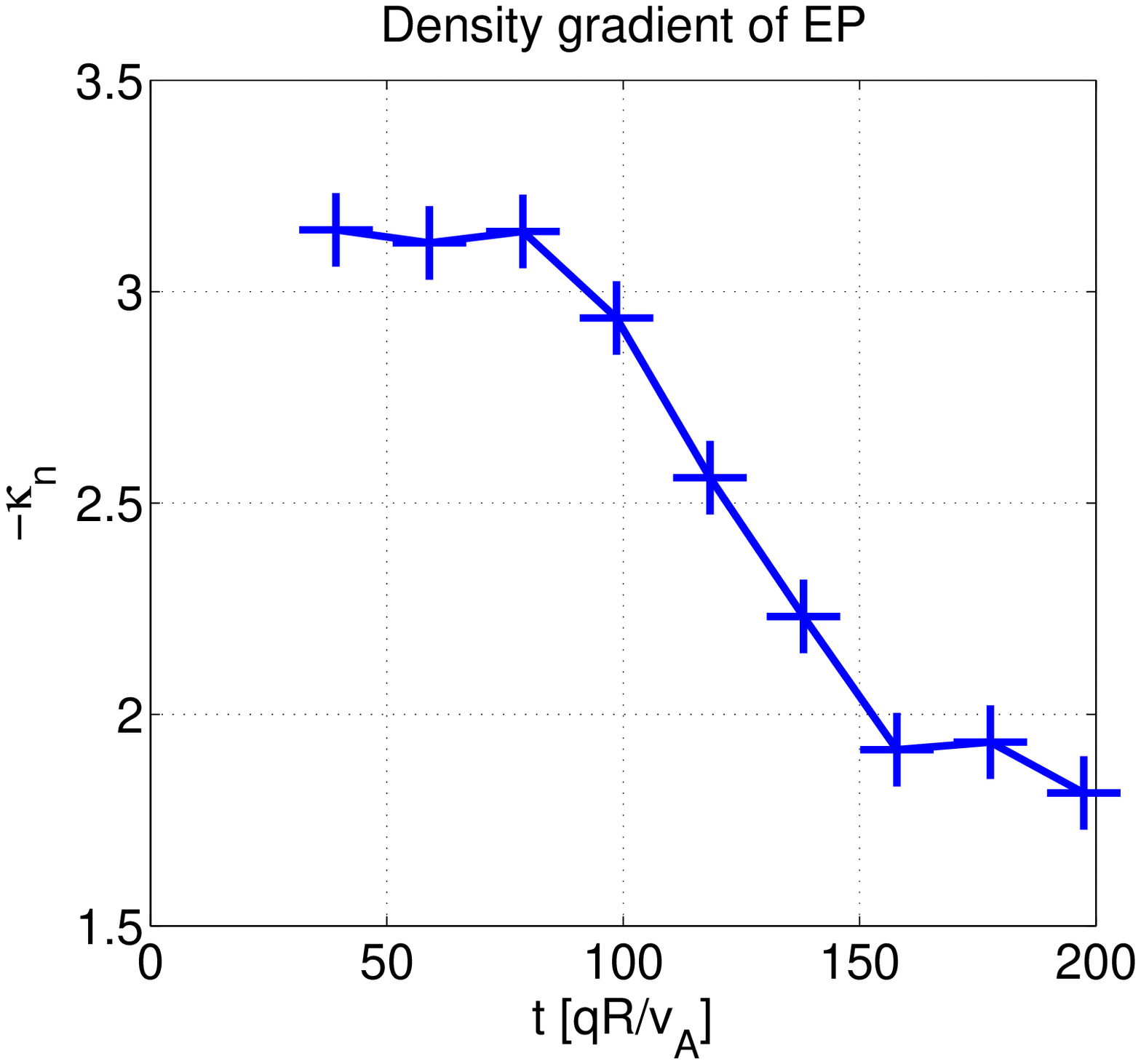}}
\vskip -1.0em
\caption{Maximum of the scalar potential measured in the poloidal plane (left) and EP density gradient at s=0.5 (right).}\label{fig:NL-phimax_t}
\end{center}
\end{figure}

\subsection{Nonlinear mode amplitude and EP profile}
\label{sec:NL-amplitude}

The evolution of the mode amplitude of a TAE and of the EP profile is described here. The EP concentration is $n_{EP}/n_e = 0.004$ and the EP temperature is $T_{EP}/T_e = 400$. The initial EP normalized gradient at s=0.5 is $\kappa_n = 3.33$. After a linear phase ($t < 100 \, qR/v_A$) the mode amplitude, measured as the maximum of the scalar potential $\phi$ in the poloidal plane, is observed to enter a ``drift-phase'', characterized by a slower subexponential growth ($100 \, qR/v_A < t < 150 \, qR/v_A$), and then a saturation ($t > 150 \, qR/v_A$) (see Fig.~\ref{fig:NL-phimax_t}). The EP profile is observed to start redistributing during the drift phase.

\subsection{Nonlinear frequency and growth rate}
\label{sec:NL-freq}

The frequency ot the TAE is observed to go up in values, starting from $\omega(t=0)\simeq 0.5 \, v_A/qR$, which is the linear value, and raising in the drift phase up to values of 
$\omega(t= 200 \, qR/v_A) \simeq 0.54 \, v_A/qR$. This value of the saturation is above the UCAP calculated according to the approximated formula of Ref.~\cite{Biancalani16PoP}, which underestimates the gap width. More generally, we can state that this saturated frequency is in the proximity of the UCAP.
The growth rate reduces drastically in the drift phase from the linear value of $\gamma(t=0)\simeq 0.08 \, v_A/qR $ to zero.

\begin{figure}[h!]
\begin{center}
\includegraphics[width=0.4\textwidth]{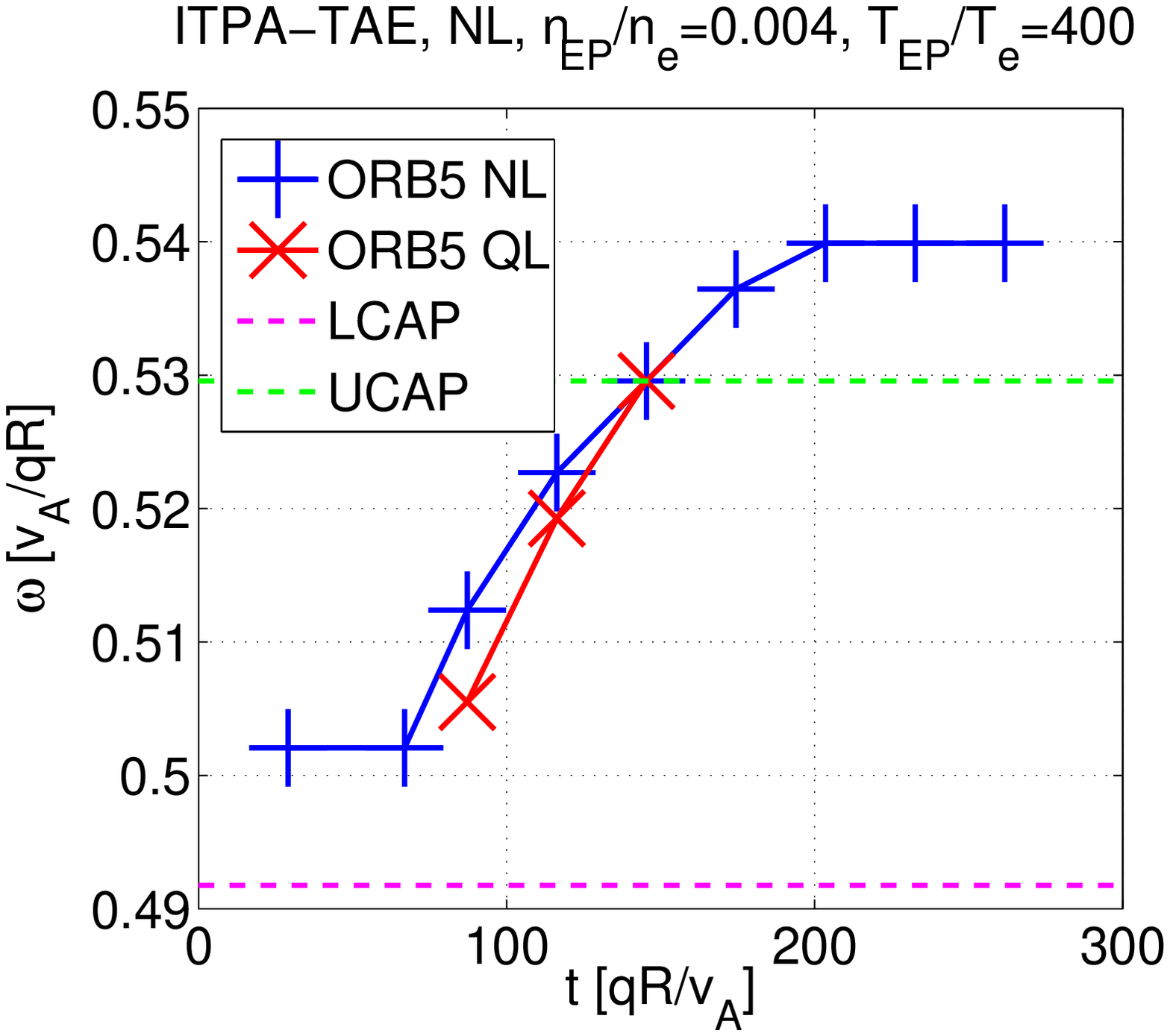}
\includegraphics[width=0.4\textwidth]{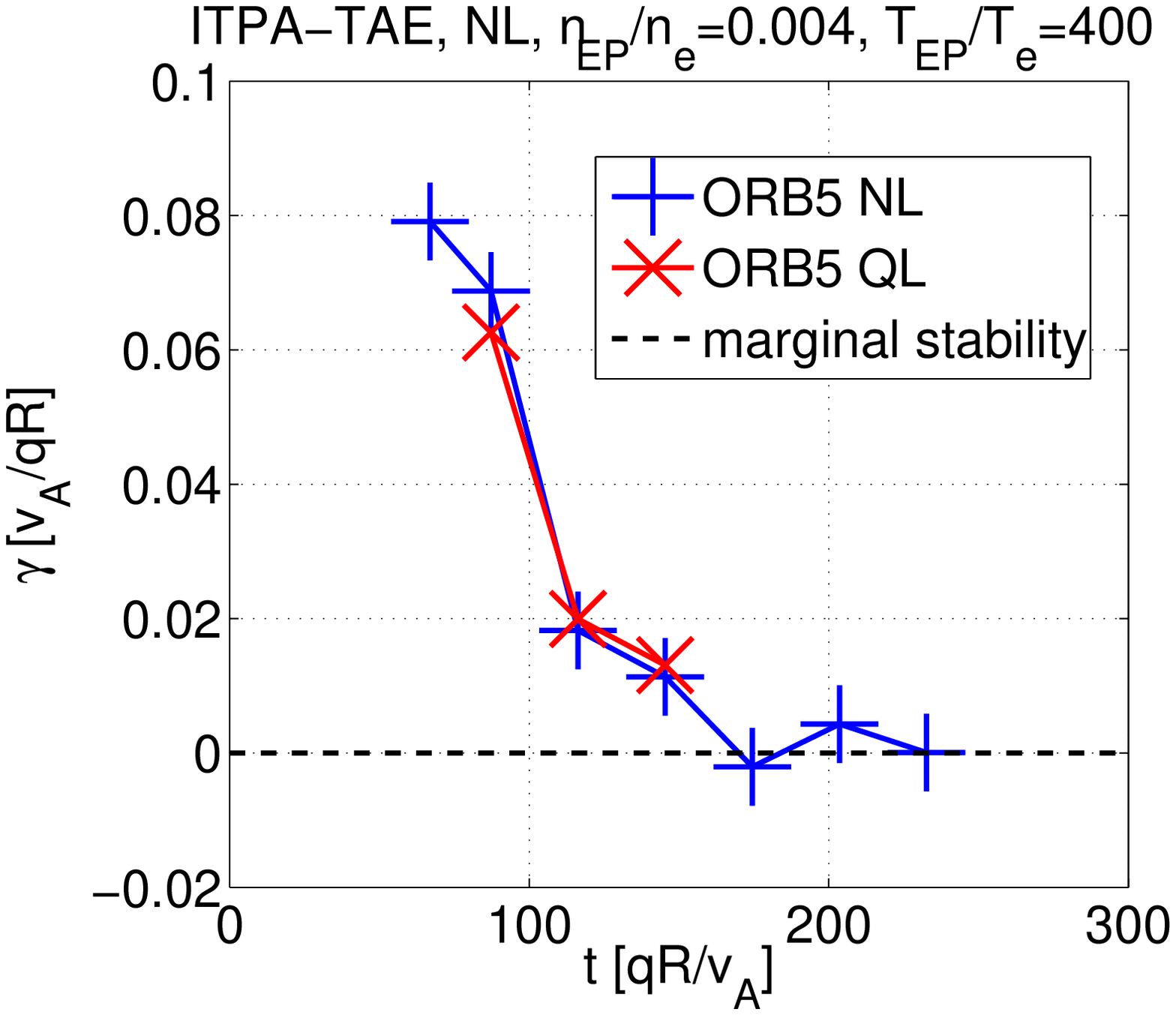}
\vskip -1.1em
\caption{Frequency (left) and growth rate (right) of a nonlinear (blue crosses) and three quasilinear simulations (red Xs, see Sec.~\ref{sec:QL}).}\label{fig:NL-QL-omegagamma}
\end{center}
\end{figure}

\begin{figure}[b!]
\begin{center}
\includegraphics[width=0.4\textwidth]{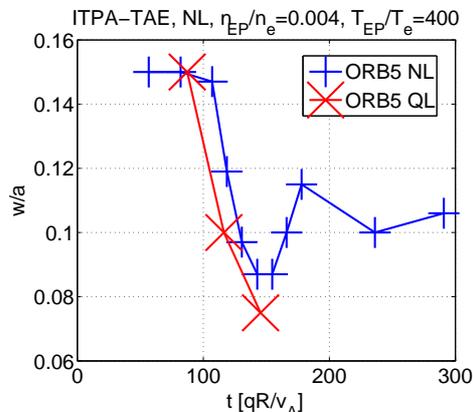}
\vskip -1.1em
\caption{Mode width of a nonlinear (blue crosses) and three  quasilinear simulations (red Xs, see Sec.~\ref{sec:QL}).}\label{fig:QL-NL-structure-2}
\end{center}
\end{figure}

\subsection{Nonlinear mode structure}
\label{sec:NL-structure-1}

During the drift phase, the radial width of the mode becomes smaller. During the saturation phase, no difference is found in the main mode structure, except for small perturbations being observed radially. 

\begin{figure}[t!]
\begin{center}
\includegraphics[width=0.48\textwidth]{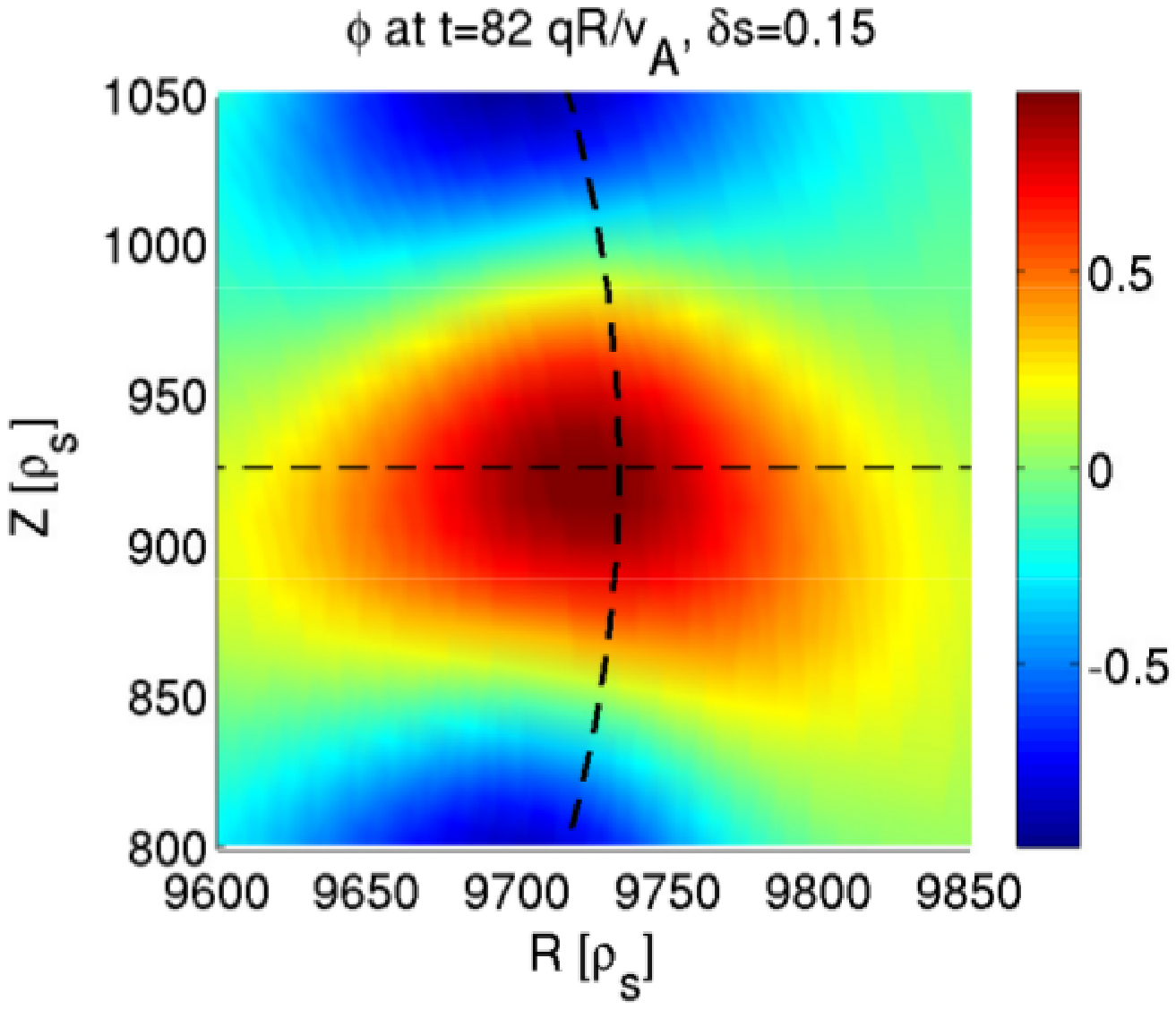}
\includegraphics[width=0.48\textwidth]{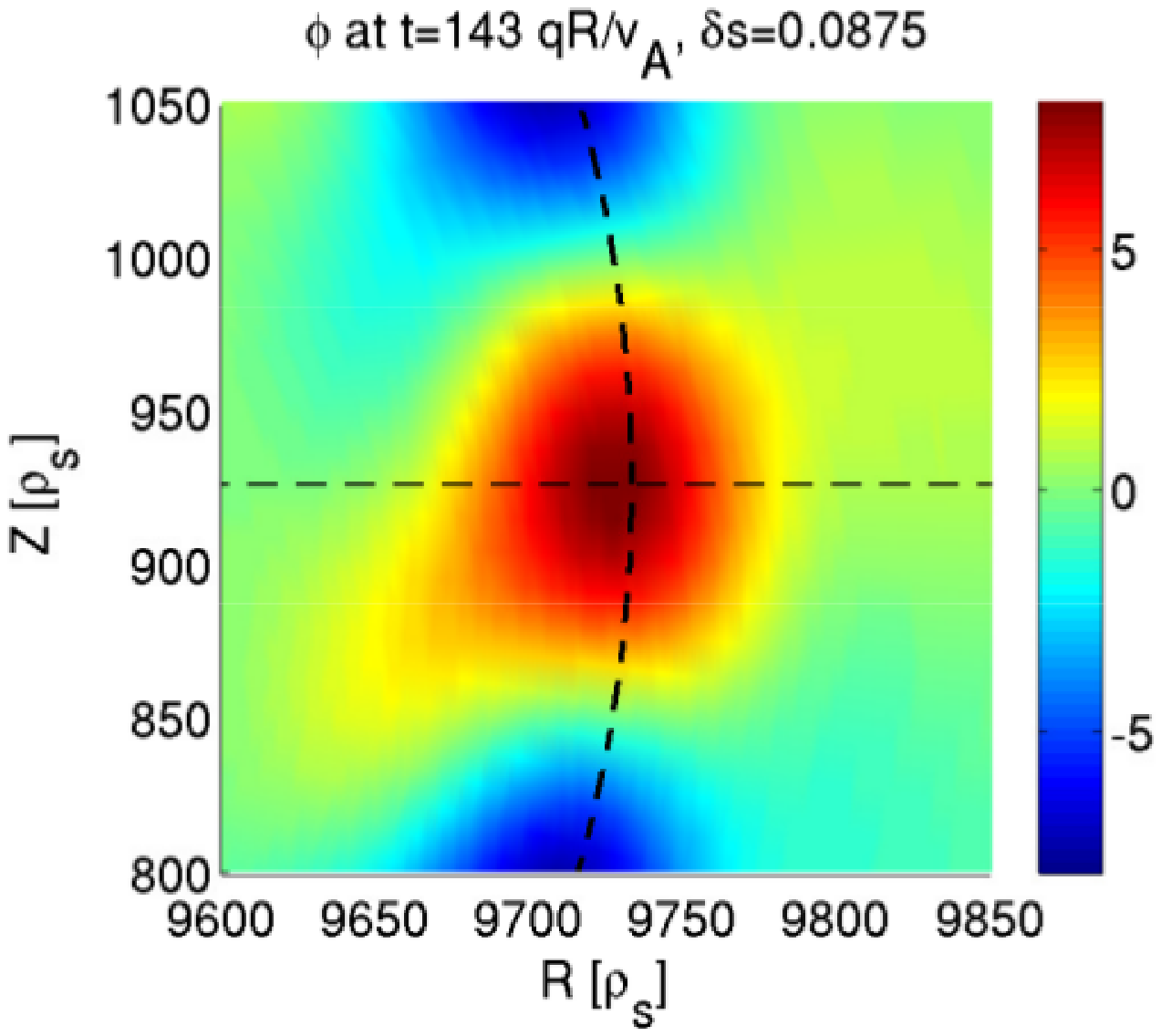}
\includegraphics[width=0.48\textwidth]{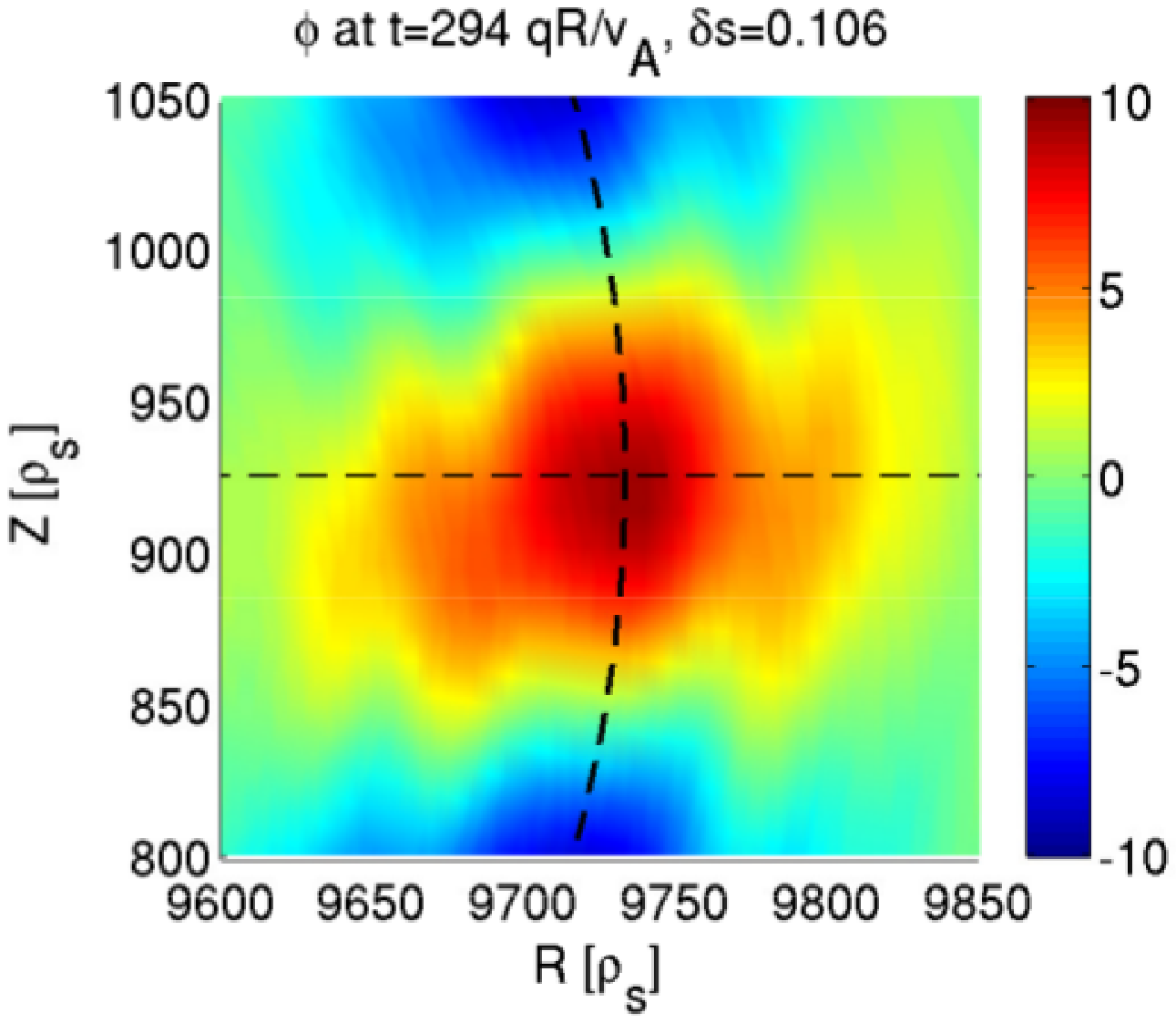}
\includegraphics[width=0.48\textwidth]{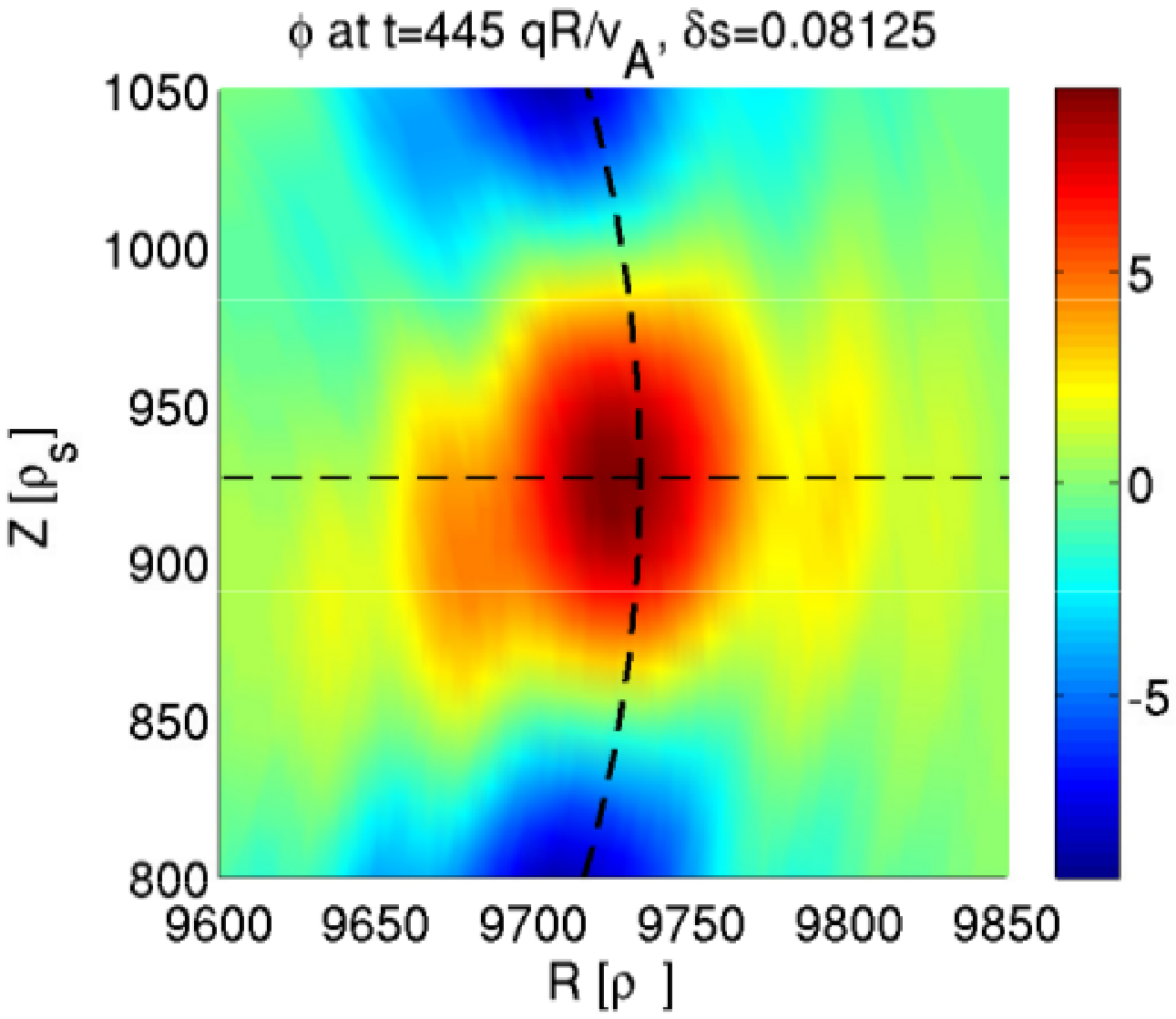}
\caption{Zoom of the scalar potential near $s=0.5, \theta=0$, depicted here at four different times, for a nonlinear simulation with $n_{EP}/n_e = 0.004$,  $T_{EP}/T_e = 400$.}\label{fig:NL-structure-1}
\end{center}
\end{figure}


Regarding modes whose frequency lies outside the continuum gap (like in the case of $T_{EP}/T_e = 200$), a different shape of the poloidal section is observed, like described in Sec.~\ref{sec:Lin-vs-TEP} for the linear phase, with a body and two wings, similarly to a ``boomerang''~\cite{Biancalani16PoP}. During the drift phase, the radial width of the boomerang body becomes smaller, the wings extension does not change, and the wings tilt. During the saturation phase, the loss of the wings generated by the EP is observed. The difference in the mode structure is accompanied by a qualitatively different mode width scaling with the linear drive, as described in Sec.~\ref{sec:NL-structure-2}.

\subsection{Dependence of nonlinear mode width on linear drive}
\label{sec:NL-structure-2}

\begin{figure}[h!]
\begin{center}
\includegraphics[width=0.45\textwidth]{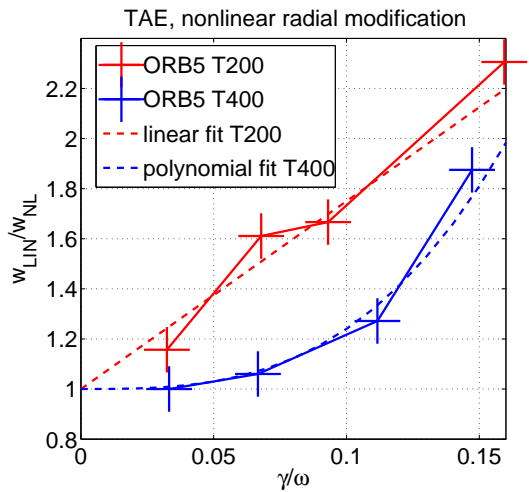}
\caption{Ratio of the linear width over the minimum  width measured in the nonlinear phase, vs the normalized growth rate, for $T_{EP}/T_e=200$ and $T_{EP}/T_e=400$.}\label{fig:NL-structure-2}
\end{center}
\end{figure}

The mode width has been found to vary during the drift phase of the NL evolution, and shown in Sec.~\ref{sec:NL-structure-1} for $n_{EP}/n_e = 0.004$,  $T_{EP}/T_e = 400$. Here, we quantify the importance of this NL structure modification, for varying intensity of the drive. The linear mode width has been qualitatively shown to increase with EP concentration (see Sec.~\ref{sec:Lin-vs-nEP}) and the quantitative comparison of the linear and NL mode width for difference EP concentration is now shown in Fig.~\ref{fig:NL-structure-2}.
The mode width here is calculated as the radial width where the amplitude of the scalar potential is half the value of the peak. At two different EP temperatures, the scaling is found to be different. In particular, for $T_{EP}/T_e=200$, the nonlinear shrinking is observed to scale linearly with the growth rate, whereas for $T_{EP}/T_e=400$, a fit with higher  power law is found.
The difference in the power law between the case at $T_{EP}/T_e=200$ and the case at $T_{EP}/T_e=400$ comes from the different nature of the two modes: the mode with $T_{EP}/T_e=200$ has frequency outside the TAE gap in the continuum (for all considered values of drives), and it has therefore the nature of an EPM, whereas the mode with $T_{EP}/T_e=400$ has frequency well within the TAE gap (for all considered values of drives) and it has therefore the nature of a pure TAE (see Fig.~\ref{fig:LIN-omegagamma_nEP}). The coefficients of the two scalings are estimated as:
\begin{equation}
w_{LIN}/w_{NL}(T_{EP}/T_e=200) \simeq  1 + 7.5 \cdot (\gamma/\omega)
\end{equation}
for the case with $T_{EP}/T_e=200$, and 
\begin{equation}
w_{LIN}/w_{NL}(T_{EP}/T_e=400) \simeq  1 + 240 \cdot (\gamma/\omega)^3
\end{equation}
for the case with $T_{EP}/T_e=400$. According to these power laws, a 10\% of nonlinear modification is found for  $\gamma/\omega\simeq 0.01$ for the case with $T_{EP}/T_e=200$, and for  $\gamma/\omega\simeq 0.07$ for the case with $T_{EP}/T_e=400$.

\subsection{Quasilinear analysis}
\label{sec:QL}
The quasilinear analysis described here consists in investigating the linear effect of the nonlinearly modified EP distribution function. With ORB5, this is done by running nonlinear simulations with small initial perturbations, and loading an instantaneous snapshot of the nonlinearly modified EP distribution function taken from the simulation depicted in Sec.~\ref{sec:NL-amplitude},~\ref{sec:NL-freq},~\ref{sec:NL-structure-1}. In this way, the linear phase can be investigated at the beginning of the nonlinear simulation.

The measured frequencies and growth rates in the quasilinear simulations are found to approximate well the frequency and growth rate evolution of the nonlinear simulation (see Fig.~\ref{fig:NL-QL-omegagamma}). This means that the EP distribution function loaded at a particular instant of the nonlinear simulation, contains all informations for determining the TAE dynamics.

The mode structure diagnosed in the quasilinear simulations shows a good match with the mode structures observed in the nonlinear simulation at different times. The radial width is also correctly reproduced with the quasilinear simulations (see Fig.~\ref{fig:QL-NL-structure-2}). 

In summary, the nonlinear dynamics of the TAE considered in this paper, and described in Sec.~\ref{sec:NL-amplitude},~\ref{sec:NL-freq},~\ref{sec:NL-structure-1}, has been reproduced with quasilinear simulations, namely simulations where the linear effects of the nonlinearly modified EP distribution function is considered. This is consistent with the fact that wave-particle nonlinearity only is acting in the nonlinear simulations described in Sec.~\ref{sec:NL-amplitude},~\ref{sec:NL-freq},~\ref{sec:NL-structure-1}.

\subsection{Theoretical interpretation}
\label{sec:theory}

\vskip 1em

The nonlinear dynamics of the reference TAE shown in Sec.~\ref{sec:NL-amplitude},~\ref{sec:NL-freq},~\ref{sec:NL-structure-1} has been described in terms of a frequency upshift and a radial shrinking occurring during the early nonlinear phase dubbed here as drift-phase. The nonlinear dynamics is in general perturbative if the following inequality is satisfied~\cite{Zonca15NJP}:
\begin{equation}
|\Delta \omega_{EP}| \ll |\Delta \omega_{SAW}| 
\end{equation}
where $\Delta\omega_{EP}$ is the nonlinear frequency modification, and $\Delta\omega_{SAW}$ is the distance of the linear mode frequency with respect to the reference CAP. For this reference case, the inequality is not satisfied, because the two terms are of the same order of magnitude (see Fig.~\ref{fig:NL-QL-omegagamma}). This confirms that, for the selected case, the EP effects are highly nonperturbative. A strong nonlinear modification of the structure, is therefore a direct consequence.

A more detailed investigation of the nonlinear evolution of the frequency can give informations also about the kind of nonlinear structure modification that should be expected. In fact, the frequency is observed to approach a CAP during the early nonlinear phase (see Fig.~\ref{fig:NL-QL-omegagamma}). By approaching the continuum, the mode which is linearly a global AE, tends nonlinearly to a singular continuum mode~\cite{Hasegawa74,Chen74}. This explains why the mode is observed to shrink radially during the early nonlinear phase. The interaction with the continuum is qualitatively different for EPMs with respect to AEs~\cite{Chen16,Chen94,Chen07}, and this yields a lower order polynomial dependence of the nonlinear frequency on the linear drive (see Fig.~\ref{fig:NL-structure-2}).

\vskip 1em

\section{Conclusions}
\label{sec:conclusions}

\vskip 1em

In this paper, an investigation of the nonlinear dynamics of the toroidicity-induced Alfv\'en Eigenmode (TAE), in the regime of the International Tokamak Physics Activity (ITPA)~\cite{Koenies12}, has been shown. The focus of this nonlinear study has been put on the wave-particle nonlinearity. The gyrokinetic particle-in-cell code ORB5, previously verified for TAEs in the linear regime~\cite{Biancalani16PoP}, has been used for this nonlinear investigation. The relevance of such study is linked to the importance of understanding the nonlinear interaction of global instabilities and energetic particle (EP) in present tokamaks and future fusion reactors.

The nonlinear analysis of the mode amplitude has shown that a TAE evolves with an initial linear phase, then a sub-exponential drift phase, and then a saturation of the mode amplitude (see also Ref.~\cite{Cole16}). In this paper, we have investigated in particular the nonlinear modification of the frequency, growth rate and mode structure. 
The frequency has been shown here to increase in time during the drift phase. The growth rate decreases during the drift phase and goes to zero when the mode saturates. The radial mode width also decreases in time. The reference TAE case has been found to fall in the highly nonperturbative regime. The radial shrinking has been explained in terms of the mode frequency approaching the continuum during the nonlinear phase. By measuring the evolution in time of the EP radial profile, we have observed that the density gradient becomes smaller in time.

A quasilinear analysis has been done, to shed light on the mechanism at the basis of the nonlinear dynamics. To this aim, we have performed nonlinear simulations with a small initial field amplitude, and with the EP distribution function loaded from particular instants of the nonlinear simulation taken as a reference. The nonlinear dynamics has been correctly reproduced with this quasilinear model. This confirms that the instantaneous dynamics in the nonlinear simulation is the linear effect of the nonlinearly modified EP distribution function, where the details in phase space are crucial.

The scaling of the nonlinear structure modification with the intensity of the linear drive has  also been described, for two particular types of modes: modes inside the continuum gap, properly labelled as TAEs, and modes outside the continuum gap, more properly referred to as energetic-particle modes (EPM). This has been achieved by considering the same equilibrium and initial mode, but driving it with two EP populations with different temperature (namely $T_{EP}/T_e = 200$ for EPMs and $T_{EP}/T_e = 400$ for TAEs), in order to have different resonance frequency. The nonlinear radial shrinking of the EPMs has been found to scale linearly with the linear drive, whereas a higher-power polynomial function is necessary to describe the nonlinear shrinking of modes in the gap. 
A nonlinear radial shrink of 10\% is found for linear growth rates above 7\% for modes with $T_{EP}/T_e = 400$, and 1\% for modes with $T_{EP}/T_e = 200$. One possible application of this result is helping defining the regime of applicability of models which aim at studying the nonlinear wave-particle interaction of Alfv\'en modes with a prescribed, fixed mode structure (see for example Ref.~\cite{Schneller16,Fitzgerald16}). A dedicated investigation of this regime of applicability, with particular interest at realistic tokamak scenarios, will be done in a dedicated paper.

\vskip 1em

\section{Acknowledgments}

\vskip 1em

Interesting discussions with E. Poli are gratefully acknowledged. This work has been carried out within the framework of the EUROfusion Consortium and has received funding from the Euratom research and training programme 2014-2018 under grant agreement No 633053, within the framework of the {\emph{Nonlinear energetic particle dynamics}} (NLED) European Enabling Research Project. The views and opinions expressed herein do not necessarily reflect those of the European Commission. Simulations were performed on the IFERC-CSC Helios supercomputer within the framework of the ORBFAST project.

\vskip 1em

\begin{appendices}

\vskip 1em

\section{Linear convergence tests}
\label{sec:appendix_conv_tests}

When decreasing the electron mass from $m_e/m_i=1/50$ to $m_e/m_i=1/2000$, we observe that in the linear phase the frequency does not sensibly change within the error bar, and the growth rate converges above $m_i/m_e=100$. The same is found for the initial nonlinear phase (the ``drift phase''), which is of interest in this paper. The deep nonlinear phase is outside the scope of this paper, and is studied elsewhere (see for example Ref.~\cite{Cole16}).

\begin{figure}[t!]
\begin{center}
\includegraphics[width=0.44\textwidth]{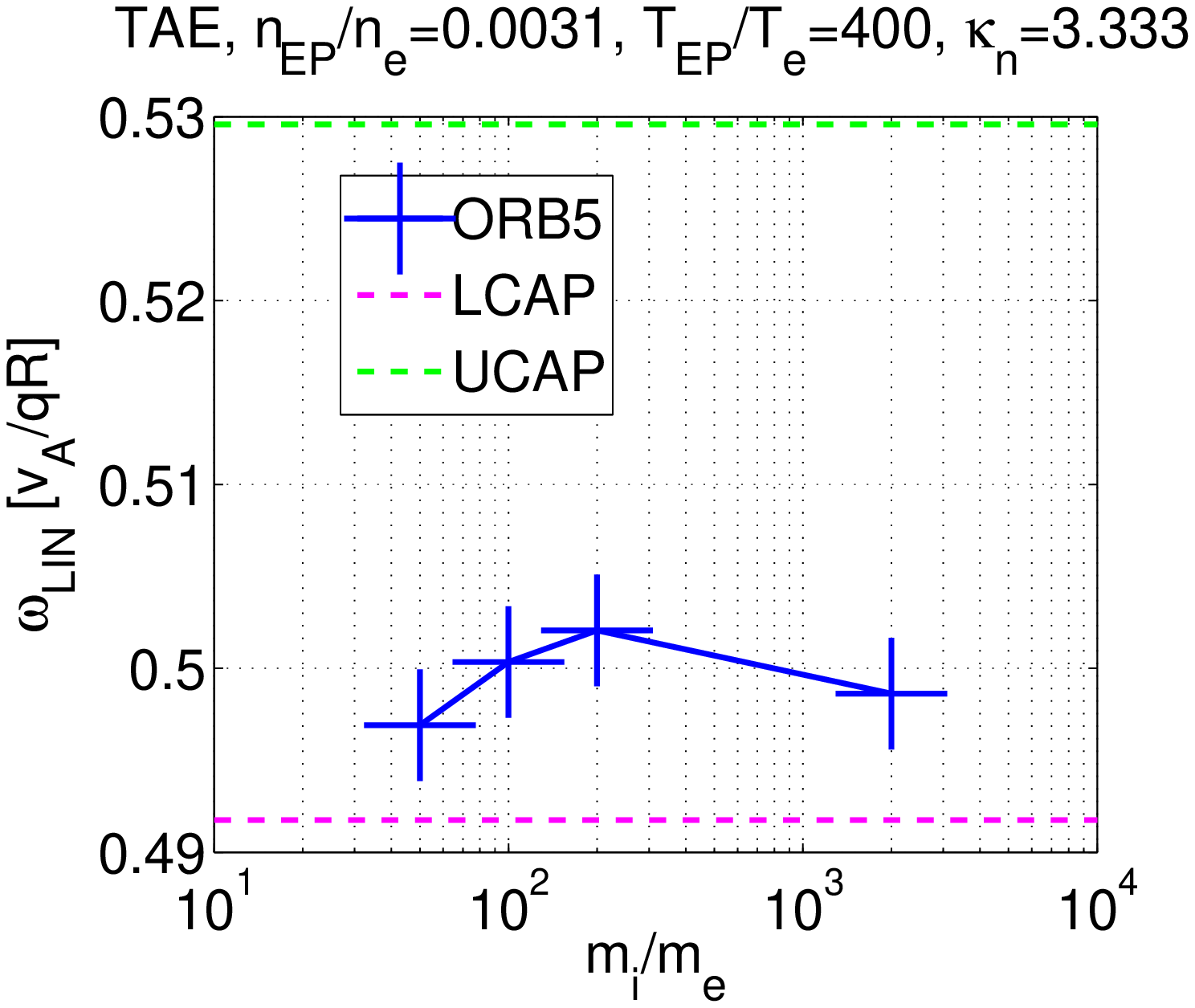}
\includegraphics[width=0.45\textwidth]{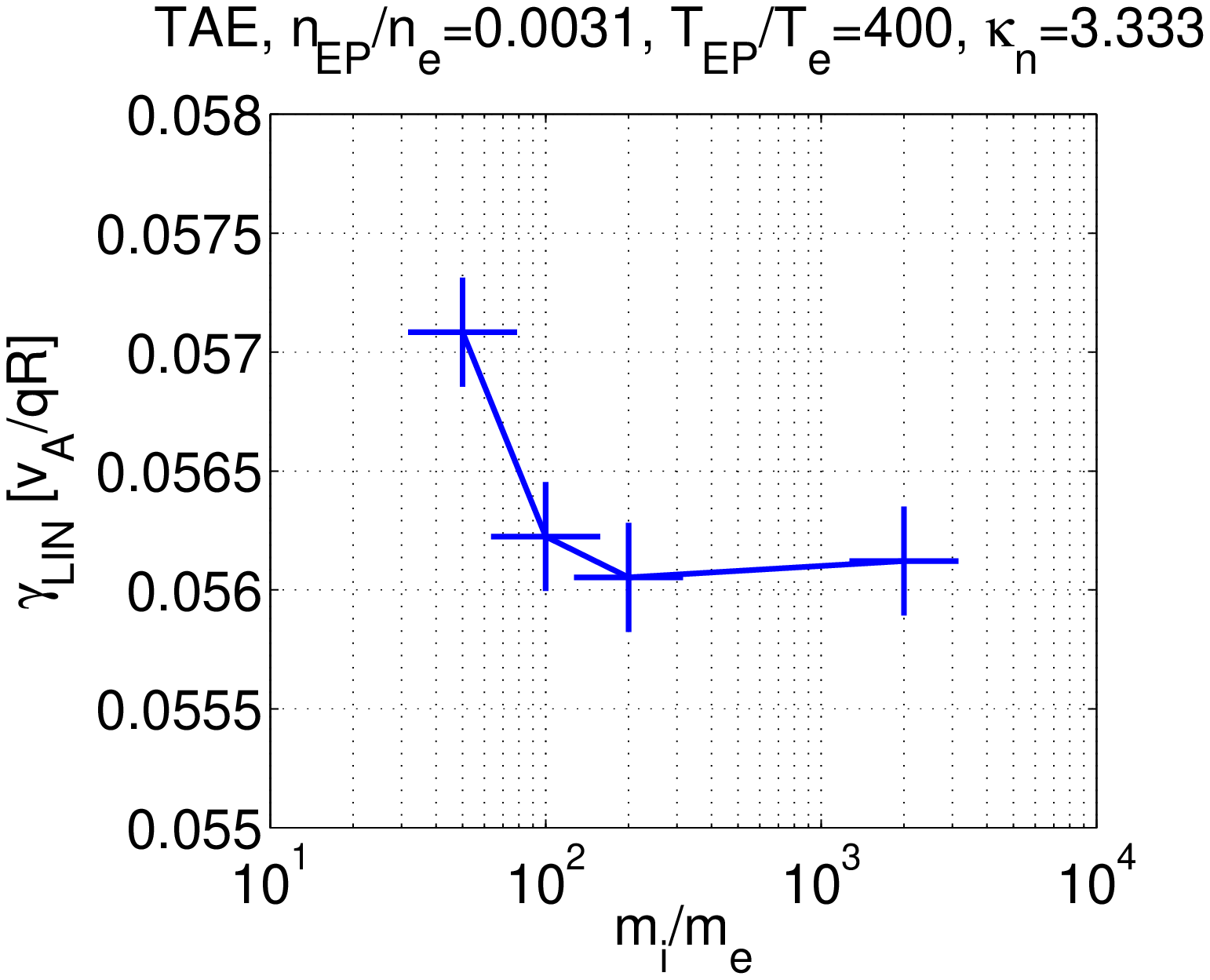}
\caption{Frequency (left) and growth rate (right) for $n_{EP}/n_e = 0.004$, $T_{EP}/T_e = 400$, and different electron masses. The frequency of the lower and upper CAPs is also shown, as dashed horizontal lines.}\label{fig:LIN-omegagamma_me}
\end{center}
\end{figure}

\newpage
\section{Linear scalings with flat i/e initial profiles}
\label{sec:appendix_flat}

\begin{figure}[b!]
\begin{center}
\includegraphics[width=0.43\textwidth]{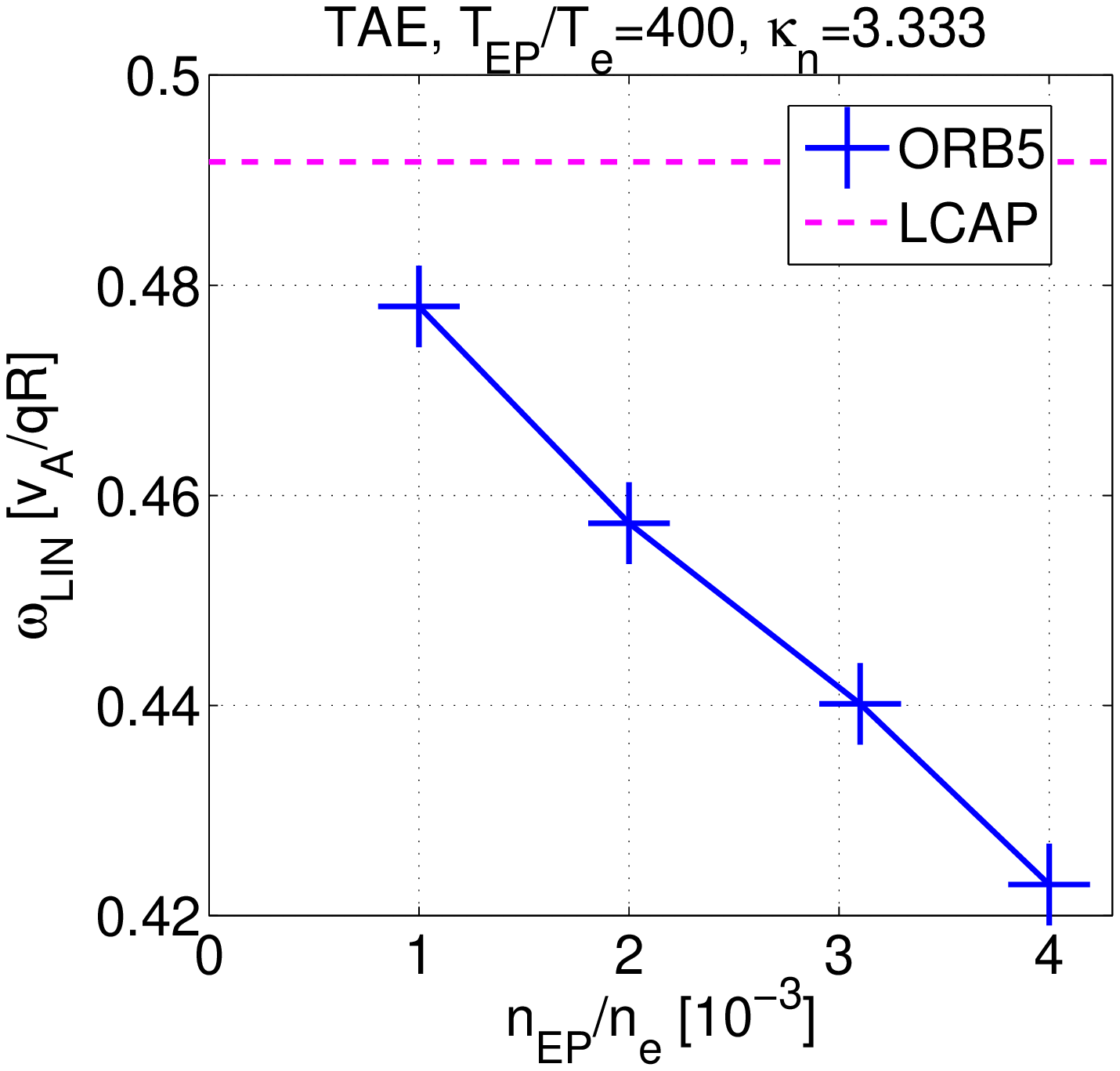}
\includegraphics[width=0.43\textwidth]{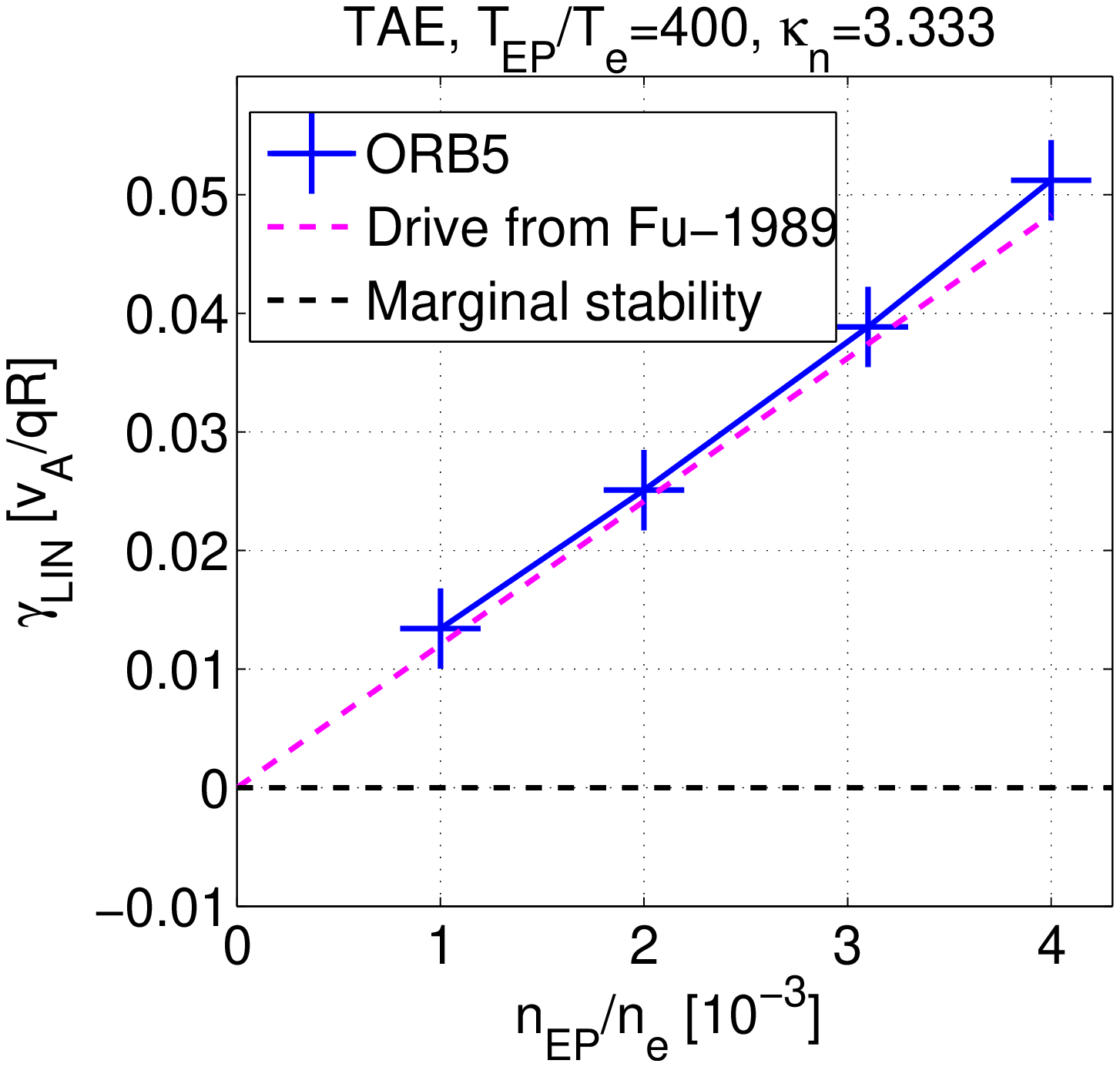}
\caption{Frequency (left) and growth rate (right) dependence on EP average density, for flat initial bulk profiles. The analytical prediction for the growth rate of Ref.~\cite{Fu89} is also shown.}\label{fig:LIN-omegagamma_nEP-FLAT}
\end{center}
\end{figure}

In order to bridge a gap with previous work~\cite{Biancalani16PoP}, we show here the scalings of frequency and growth rates, when the equilibrium profiles of the bulk ions and electrons are initialized as flat. The quasineutrality here is not imposed at t=0, but it is achieved automatically by the code which redistributes the profiles in the first time steps. In this case, with increasing EP concentration, we observe a linearly decresing value of frequency (entering the continuum below the LCAP) and a linearly increasing value of growth rate. This is a regime closer to that considered in Ref.~\cite{Fu89}, whose analytical prediction of the growth rate gives a good match with the scaling obtained numerically (see Fig.~\ref{fig:LIN-omegagamma_nEP-FLAT}, where the theoretical damping is neglected).

\begin{figure}[t!]
\begin{center}
\includegraphics[width=0.44\textwidth]{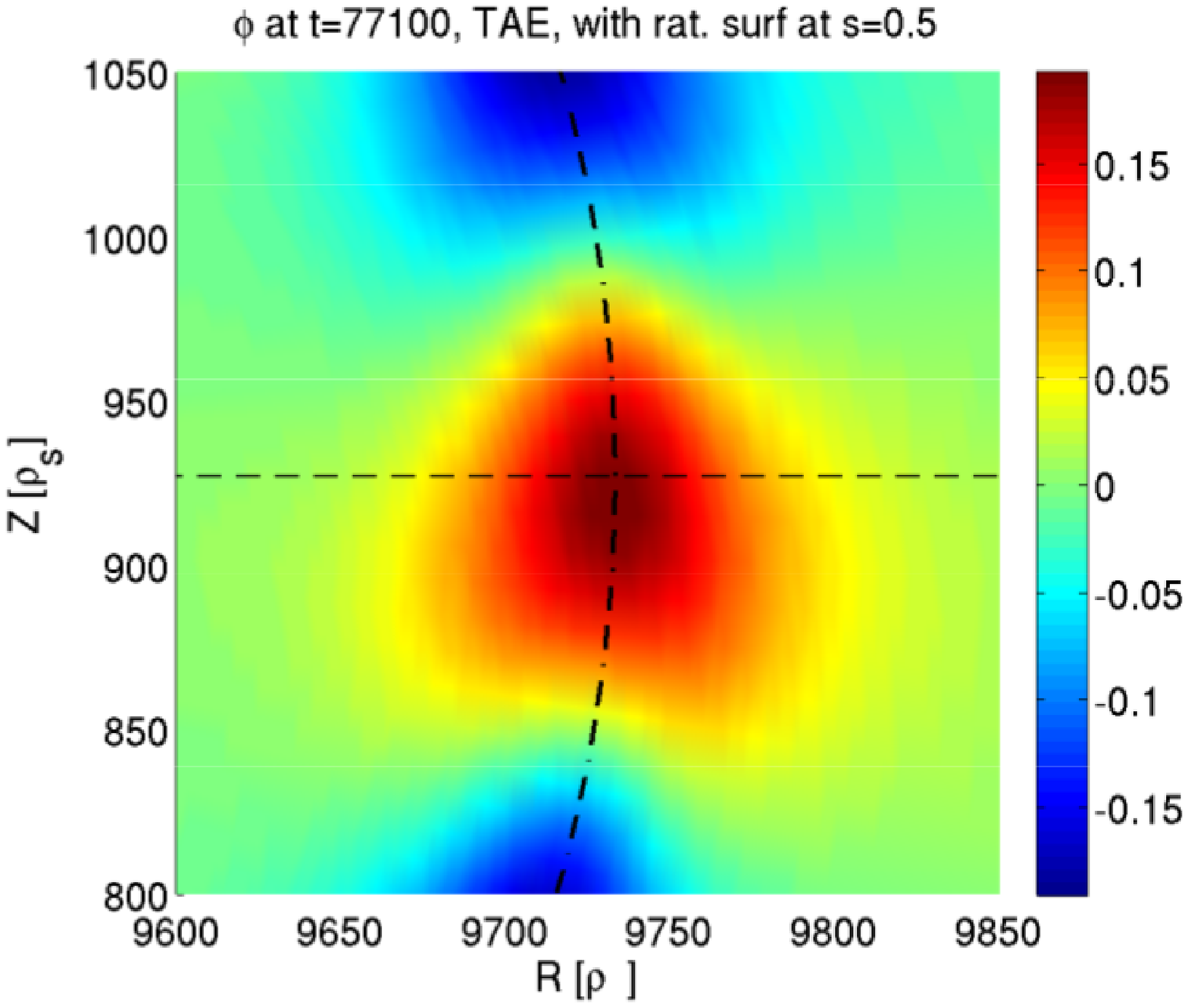}
\includegraphics[width=0.45\textwidth]{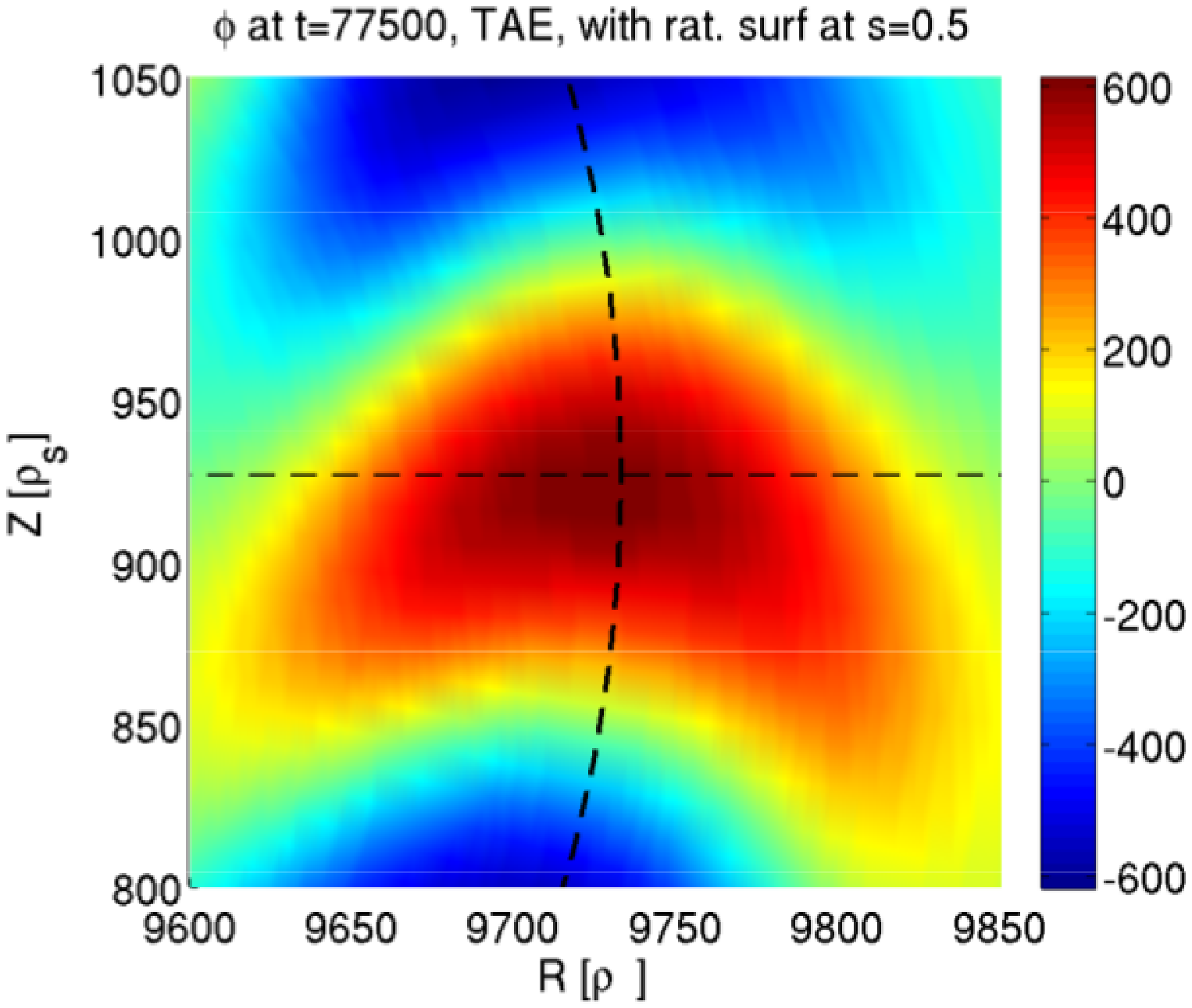}
\caption{Structure in the poloidal plane, for $n_{EP}/n_e = 0.001$ (left) and $n_{EP}/n_e = 0.004$ (right). Here $T_{EP}/T_e = 400$, and $\kappa_n = 3.333$.}\label{fig:LIN-structure_nEP-FLAT}
\end{center}
\end{figure}
Due to the low frequency, with respect to the LCAP, the structure of the TAE  shows an evident ``boomerang''~\cite{Biancalani16PoP} (or ``croissant'') shape.  When increasing the EP concentration, we observe that both the radial width of the body and the wings extension increase (see Fig.~\ref{fig:LIN-structure_nEP-FLAT}).

\end{appendices}

\end{document}